\documentclass[11pt, a4paper]{article}
\usepackage[T1]{fontenc}
\usepackage{a4wide, url}
\usepackage[english]{babel}
\usepackage{graphicx}
\usepackage{array}
\usepackage{natbib}
\usepackage[textfont=sc,labelfont=bf]{caption}
\usepackage{setspace}
\usepackage{multirow}
\usepackage{rotate}
\usepackage{amsthm}
\usepackage{amsmath}
\usepackage{amsfonts}
\usepackage{amssymb}
\usepackage{pgf} 
\usepackage{bm}
\usepackage{paralist}
\usepackage{pdflscape}
\usepackage{xfrac}
\usepackage[linesnumbered,ruled,vlined]{algorithm2e}
\SetKwInput{KwInput}{Input}                
\SetKwInput{KwOutput}{Output}              
\usepackage{enumitem,kantlipsum}

\usepackage{comment}
\excludecomment{ext}\excludecomment{comment}

\usepackage{tabu}

\newcolumntype{L}[1]{>{\raggedright\let\newline\\\arraybackslash\hspace{0pt}}m{#1}}
\newcolumntype{C}[1]{>{\centering\let\newline\\\arraybackslash\hspace{0pt}}m{#1}}
\newcolumntype{R}[1]{>{\raggedleft\let\newline\\\arraybackslash\hspace{0pt}}m{#1}}

\DeclareMathAlphabet\mathbfcal{OMS}{cmsy}{b}{n}

\newcommand{\mbf}{\mathbf}
\newcommand{\beq}{\begin{equation}}
\newcommand{\eeq}{\end{equation}}
\newcommand{\bea}{\begin{eqnarray}}
\newcommand{\eea}{\end{eqnarray}}
\newcommand{\ba}{\begin{array}}
\newcommand{\ea}{\end{array}}
\newcommand{\bit}{\begin{itemize}}
\newcommand{\eit}{\end{itemize}}
\newcommand{\ben}{\begin{enumerate}} 
\newcommand{\een}{\end{enumerate}}
\newcommand{\bpm}{\begin{pmatrix}}
\newcommand{\epm}{\end{pmatrix}}
\newcommand{\bbm}{\begin{bmatrix}}
\newcommand{\ebm}{\end{bmatrix}}

\renewcommand{\l}{\left}
\renewcommand{\r}{\right}

\newcommand{\E}[0]{\mathrm{E}}
\newcommand{\Var}[0]{\mathrm{Var}}

\newcommand{\nn}{\nonumber}
\newcommand{\wh}{\widehat}
\newcommand{\wt}{\widetilde}

\newtheorem{ass}{Assumption}

\newtheorem{prop}{Proposition}

\title{\textsc{\large 	
Quasi Maximum Likelihood Estimation of Non-Stationary Large Approximate Dynamic Factor Models}}
\date{ }

\begin{document}
\maketitle

\begin{center}\vspace{-1.5cm}
\begin{tabular}{cp{1cm}c}
 Matteo Barigozzi &&  Matteo Luciani\\[.1cm] 
\footnotesize London School of Economics &&\footnotesize Federal Reserve Board\\[.1cm] 
\footnotesize  m.barigozzi@lse.ac.uk &&\footnotesize matteo.luciani@frb.gov\\[1cm] 
\end{tabular}

\small \today\\[1.5cm]
\end{center}

\begin{abstract}
This paper considers estimation of large dynamic factor models with common and idiosyncratic trends by means of the Expectation Maximization algorithm, implemented jointly with the Kalman smoother. We show that, as the cross-sectional dimension $n$ and the sample size $T$ diverge to infinity, the common component for a given unit estimated at a given point in time is $\min(\sqrt n,\sqrt T)$-consistent. The case of local levels and/or local linear trends trends is also considered. By means of a MonteCarlo simulation exercise, we compare our approach with estimators based on principal component analysis.

\vspace{0.5cm}


\end{abstract}

\renewcommand{\thefootnote}{$\ast$} 
\thispagestyle{empty}

\footnotetext{
A preliminary version of the results in this paper was made available with the title ``Common factors, trends, and cycles in large datasets'' (2017), by M. Barigozzi and M. Luciani, arXiv:1709.01445, and 2017-111, Board of Governors of the Federal Reserve System.\\

Disclaimer: the views expressed in this paper are those of the authors and do not necessarily reflect the views and policies of the Board of Governors or the Federal Reserve System.  } 

\renewcommand{\thefootnote}{\arabic{footnote}}

%
%
\newpage
\section{Introduction}
In the last fifteen years, large dimensional stationary factor models have achieved great success in the economic profession, especially in forecasting macroeconomic variables \citep[see, e.g.,][]{Nowcasting}, and are now a common tool in several policy institutions. However, macroeconomic time series are typically non-stationary due to the presence of common and idiosyncratic stochastic trends, and the practice of differencing the data to achieve stationarity is a problem that not always has a clear-cut solution. Take for example the case of the unemployment rate, which is a highly-persistent time series, but at the same time economic theory forbids it to have a unit root; or, take as another example the case of inflation, which shows  periods of high-persistence in the late 70s early 80s, while more recently displays clear mean reversion. To avoid the risk of over- or under-differencing data, a Non-Stationary Dynamic Factor Model (NS-DFM) is then desirable, and it is studied in this paper. 

The NS-DFM proposed in this paper captures several features of macroeconomic data as it takes into account the presence of common trends generating permanent fluctuations in the economy, as well as common transitory forces generating cyclical fluctuations. More technically, in our model, the common factors are a cointegrated vector process, thus containing both $I(1)$ trends and stationary components. Moreover, the NS-DFM addresses the possible presence of idiosyncratic trends, as well as the presence of secular (linear) trends, which can have either a constant slope (deterministic linear trends) or a time-varying slope (local linear trends). 

In this paper, we study estimation of the NS-DFM by Quasi Maximum Likelihood (QML) implemented through the Expectation Maximization (EM) algorithm and the Kalman smoother (KS). Specifically, we extend the results in \citet{BLqml} for the stationary case, to prove that when the common factors are the only source of non-stationarity, the common component estimated at a given point in time and for a given unit is $\min(\sqrt n,\sqrt T)$-consistent. We also discuss extensions to the cases of (i) unit roots in the idiosyncratic components, and (ii) local levels and local linear trends. 
 
Estimation is implemented in two steps. First, given the observed data, by means of the KS we estimate the conditional mean of the latent factors,  which, together with its associated conditional covariance matrix, we use to compute the expected log-likelihood of the model (E-step).\footnote{In a non-stationary setting the existence of the conditional mean of the factor as a minimizer of the mean-squared prediction error has been proved by \citet[Theorem 1]{hannan67} and \citet[Theorem 1]{sobel67}.} Second, we maximize the expected log-likelihood with respect to the loadings and the other parameters of the model (M-step). The use of an iterative procedure to extract unobserved components in the case of non-stationary data was proposed since the original work by \citet{kalman60}. Although this is not the first paper using these techniques for non stationary data, this is the first paper to address consistency of factors. Moreover, QML estimation of autoregressive processes with unit roots is a classical problem studied at length by the literature \citep{simsstockwatson,johansen91}.\footnote{Solutions based on spectral analysis are also in \citet{bell84} and \citet{CT76}.}

Estimation of NS-DFMs has also been studied by \citet{baing04} and \citet{BLL2} by PC analysis on differenced data. Both approaches are designed to account for non-stationary idiosyncratic components; however, only the latter is designed to deal with linear deterministic trends. \citet{bai04} has used a factor model to estimate common trends via PC on data in levels. However, because of its nature, that approach is valid only if all idiosyncratic components are stationary, i.e., only if data are cointegrated.

Compared to those PC based estimators, our approach has a number of practical advantages. First, it allows estimating the model even in the presence of missing values, which is crucial when using the model in real-time because macroeconomic data are published with delays and at non-synchronized dates. Second, it allows estimating jointly stochastic trends as well as (deterministic or local) linear trends, whereas \citet{baing04} and \citet{BLL2} are forced to remove the deterministic trends before running PC analysis. Third, it allows having time-varying parameters, such as, for example, the slope of linear trends. Fourth, it allows putting restrictions on the parameters, such as national accounts identities, or restrictions coming from economic theory.

From a theoretical point of view, our estimator converges at a faster rate than those of \citet{baing04} and \citet{BLL2}. However, this faster convergence does not come for free. Indeed, our estimator is based on stronger assumptions than those of PC analysis: namely, it is derived under the assumption that we know which idiosyncratic components are $I(1)$ and which ones are stationary, and which series have a linear trend component. Under this assumption, we can model the $I(1)$ idiosyncratic components, and the time-varying slopes or means, as additional latent states in the KS, thus allowing to simultaneously estimate the entire model. This strategy is shown to work well in practice, provided the number of additional latent states is not too large.

The use of the EM in time series dates back to \citet{SS77}, \citet{shumwaystoffer82}, \citet{watsonengle83}, \citet{quahsargent93}, and \citet{SAZ13}, among others. However, with the exception of the last two, none of the above works has considered the case of non-stationary data. Moreover, to the best of our knowledge, no asymptotic theory exists for the setting considered in this paper.

The rest of the paper proceeds as follows: in Section \ref{sec:modelNS}, we present the NS-DFM and its assumptions. Estimation is outlined in Section \ref{sec:estNS} where we also prove consistency. The extension to non-stationary idiosyncratic states is discussed in Section \ref{sec:idioI1}. Numerical results are in Section \ref{sec:mc2}. Section \ref{sec:conclusion} concludes.

%
%
\section{Model and assumptions}\label{sec:modelNS}
We define a NS-DFM driven by $q$ factors as 
\begin{align}
x_{it}&=\alpha_{it}+\beta_{it} t+\bm b_i^\prime(L) \bm f_t +\xi_{it},\label{eq:NSDFM1}\\
\bm f_t&= \bm{\mathcal A}(L)\bm f_{t-1}+\bm u_t,\label{eq:NSDFM2}\\
\xi_{it}&=\rho_i\xi_{it-1}+e_{it},\label{eq:NSDFM3}\\
\alpha_{it}&=\alpha_{it-1}+\omega_{it},\label{eq:NSDFM5}\\
\beta_{it}&=\beta_{it-1}+\eta_{it},\label{eq:NSDFM4}
\end{align}
for  $i=1,\ldots, n$, and $t=1,\ldots, T$. We let $\chi_{it}=\bm b_i^\prime(L) \bm f_t$. Then, $\bm\chi_{nt}=(\chi_{1t}\cdots\chi_{nt})^\prime$ is the common component, $\bm\xi_{nt}=(\xi_{1t}\cdots\xi_{nt})^\prime$ the idiosyncratic component, $\bm {\mathcal B}_n(L)=(\bm b_1(L)\cdots\bm b_n(L))^\prime$ the $n\times q$ polynomial matrix of factor loadings, $\bm f_t=(f_{1t}\cdots f_{qt})^\prime$ the $q$ factors, $\bm u_t=(u_{1t}\cdots u_{qt})^\prime$ the $q$ common shocks, $\mbf e_{nt}=(e_{1t}\cdots e_{nt})^\prime$ the idiosyncratic shocks, and we also define $\bm\omega_{nt}=(\omega_{1t}\cdots\omega_{nt})^\prime$ and $\bm\eta_{nt}=(\eta_{1t}\cdots\eta_{nt})^\prime$.

We make the following assumptions.

\begin{ass}\label{ass:dynamic}
\begin{inparaenum}[(a)]
\item for all $i\in\mathbb N$ and $z\in\mathbb C$, $\bm b_i(z)=\sum_{k=0}^s \bm b_{ik}z^k$, such that $\bm b_{ik}$ are $q\times 1$ and $s$ is a finite integer with $s\ge 0$;
\item for all $n\in\mathbb N$, let $\bm{\mathcal B}_{kn}=(\bm b_{1k}\cdots \bm b_{nk})^\prime$, then $\lim_{n\to\infty}\Vert n^{-1}\bm{\mathcal B}_{kn}^\prime\bm{\mathcal B}_{kn}-\bm\Sigma_{k}\Vert=0$, with $\bm\Sigma_{k}$ being $q\times q$, and $\bm \Sigma_0$ positive definite, while $\mbox{rk}(\bm\Sigma_k)\le q$ for $k=1,\ldots,s$, 
moreover, for all $i\in\mathbb N$ and $k=0,\ldots, s$, $\Vert\bm b_{ik}\Vert \le M_B$ for some finite positive real $M_B$ independent of $i$ and $k$; \item $\bm\Gamma^{\Delta f}=\E_{\varphi_n}[\Delta\bm f_t\Delta\bm f_t^\prime]$ is $q\times q$ positive definite and there exists a finite positive real $M_f$, such that $\Vert\bm\Gamma^{\Delta f}\Vert\le M_f$;
\item $q$ is a finite positive integer, such that $q<n$ and is independent of $n$; \item $\bm{\mathcal A}(z)=\sum_{k=1}^{p} \bm{\mathcal A}_{k}z^{k-1}$, such that $\bm{\mathcal A}_{k}$ are $q\times q$ and $p$ is a finite positive integer, and $\det(\mbf I_q-\bm{\mathcal A}(z))\ne 0$ for all $z\in\mathbb C$ such that $|z|< 1$; \item $\mbox{\upshape rk}(\bm{\mathcal A}(1))=d$ with $0<d\le q$; \item $|\rho_i|\le 1$ for all $i\in\mathbb N$; 
\item $\alpha_{i0}$ and $\beta_{i0}$ are finite reals.
\end{inparaenum}
\end{ass}

\begin{ass}
\begin{inparaenum}[(a)]\label{ass:modelNS}
 \item for all $t\in\mathbb Z$, $\bm u_t\sim \mathcal N(\mbf 0_q,\bm\Gamma^u)$, such that $\bm\Gamma^u$ is $q\times q$ and positive definite, and  $\E_{\varphi_n}[\bm u_{t}\bm u_{t-k}^\prime]=\mbf 0_{q\times q}$ for all $k\ne 0$; \item for all $t\in\mathbb Z$ and all $n\in\mathbb N$, $\mbf e_{nt}\sim\mathcal N(\mbf 0_n,\bm\Gamma_n^e)$, such that $\bm\Gamma_n^e$ is $n\times n$ and positive definite, and $\E_{\varphi_n}[\mbf e_{nt}\mbf e_{nt-k}]=\mbf 0_{n\times n}$ for all $k\ne 0$; 
 \item for all $n\in\mathbb N$, $\Vert\bm\Gamma_n^e \Vert \le M_e$, for some positive real $M_e$ independent of $n$; \item $\E_{\varphi_n}[\mbf e_{nt}\bm u_{s}^\prime]=\mbf 0_{n\times q}$ for all $n\in\mathbb N$ and $t,s\in\mathbb Z$; \item for all $t\in\mathbb Z$ and all $n\in\mathbb N$, $\bm \omega_{nt}\sim\mathcal N(\mbf 0_n,\bm\Gamma_n^\omega)$ and $\bm \eta_{nt}\sim\mathcal N(\mbf 0_n,\bm\Gamma_n^\eta)$, such that $\bm\Gamma_n^\omega$ and $\bm\Gamma_n^\eta$ are diagonal, respectively with entries $0\le \sigma_{i\omega}^{2} < M_\omega$ and $0\le \sigma_{i\eta}^{2} < M_\eta$, for some positive reals $M_\omega$ and $M_\eta$ independent of $i$, and $\E_{\varphi_n}[\bm \omega_{nt}\bm \omega_{nt-k}]=\mbf 0_{n\times n}$ and $\E_{\varphi_n}[\bm \eta_{nt}\bm \eta_{nt-k}]=\mbf 0_{n\times n}$ for all $k\ne 0$; \item 
 $\E_{\varphi_n}[\bm\omega_{nt}\bm u_{s}^\prime]=\mbf 0_{n\times q}$,
 $\E_{\varphi_n}[\bm\eta_{nt}\bm u_{s}^\prime]=\mbf 0_{n\times q}$, for all $n\in\mathbb N$ and $t,s\in\mathbb Z$;
 \item  $\E_{\varphi_n}[\bm\omega_{nt}\mbf e_{ns}^\prime]=\mbf 0_{n\times n}$,
 $\E_{\varphi_n}[\bm\eta_{nt}\mbf e_{ns}^\prime]=\mbf 0_{n\times n}$, and
 $\E_{\varphi_n}[\bm\omega_{nt}\bm \eta_{ns}^\prime]=\mbf 0_{n\times n}$,
 for all $n\in\mathbb N$ and $t,s\in\mathbb Z$.
\end{inparaenum}
\end{ass}

\begin{ass}\label{ass:I1}
For any given $n\in\mathbb N$, there exists sets $\mathcal I_1\in\{1,\ldots,n\}$, $\mathcal I_a\in\{1,\ldots,n\}$,  and $\mathcal I_b\in\{1,\ldots,n\}$, such that:
\begin{inparaenum}[(a)]
\item if $i\in\mathcal I_1$ then $\rho_i=1$, while $\rho_i=0$ otherwise, moreover $\# \mathcal I_1=n_1$ such that $n_1n^{-1}\to 0$, as $n\to\infty$;
\item if $i\in\mathcal I_a$ then $\sigma_i^{2\omega}\ge C_\omega$ for some positive real $C_\omega$, while $\sigma_i^{2\omega}=0$ otherwise, moreover $\# \mathcal I_a=n_a$ such that $n_an^{-1}\to 0$, as $n\to\infty$;
\item if $i\in\mathcal I_b$ then $\sigma_i^{2\eta}\ge C_\eta$ for some positive real $C_\eta$, while $\sigma_i^{2\eta}=0$ otherwise, moreover $\# \mathcal I_b=n_b$ such that $n_bn^{-1}\to 0$, as $n\to\infty$.
\end{inparaenum}
\end{ass}

By Assumption \ref{ass:dynamic}(a) we are considering the case in which factors are loaded dynamically with a finite number of lags. We do not consider here the case of autoregressive filters, which has been studied in \citet{FHLZ17} in the stationary case. By Assumption \ref{ass:dynamic}(b) we are assuming for simplicity that all $q$ factors are pervasive at lag-zero, while at higher lags they might or might not have a pervasive effect depending on the rank of $\bm\Sigma_k$. In other words, \eqref{eq:NSDFM1} can be seen as a factor model with $q(s+1)$ factors of which $q$ are strong factors, i.e., having an effect on all series, and the remaining $qs$ are weak factors, i.e., having an effect only on a subset of series. 

By Assumptions \ref{ass:dynamic}(d) and \ref{ass:dynamic}(e), when $d<q$ we allow the dynamics of the factors to be driven by $(q-d)<q$ unit roots implying the presence of $(q-d)$ common trends \citep{stockwatson88JASA}. Clearly, in this setting, the factors are cointegrated with cointegration rank $d$, thus representing the permanent and transitory aspects of macroeconomic dynamics. When $d=0$---i.e., the dynamics of the factors are driven by $q$ unit roots---the VAR for the common factors in levels in \eqref{eq:NSDFM2} does not exist; instead, it exists a VAR for $\Delta \bm f_t$, or the factors can be modeled as $q$ independent random walks as in \citet{bai04}. That said, the case $d>0$ is the relevant one, as there is full agreement in the economic profession that while some fluctuations in the economy are permanent (common trends), some others are only temporary.

Assumption \ref{ass:modelNS} characterizes the innovations of the model. In particular, by part (c)  the idiosyncratic innovations $e_{it}$ are allowed to be mildly cross-correlated, thus implying that $\Delta x_{it}$ follows an approximate factor model. Moreover, by part (e) we allow some series to be driven by a time-varying intercept and/or a trend with time-varying slope, modeled as in a local level and local linear trend model, respectively \citep[Section 2.3.6, page 45]{harvey90}. Notice that, if we set $\sigma_{i\eta}^2=0$, then the trend becomes deterministic with slope $\beta_{i0}$, which is fixed to a constant by Assumption \ref{ass:modelNS}(h), and similarly if we set $\sigma_{i\omega}^2=0$, we have a deterministic, hence constant, intercept term equal to $\alpha_{i0}$. Finally, by parts (d) and (f) all innovations are independent. Notice that gaussianity is not strictly needed, but it is a reasonable assumption in macroeconomics.

Under these assumptions, it can be shown that the covariance matrix of the differenced common component $\Delta\bm\chi_{nt}$ has at least $q$ and at most $q(s+1)$ eigenvalues that diverge linearly as $n\to\infty$. In particular, letting the covariance matrix of $\Delta \bm \chi_n$ be $\bm\Gamma_n^{\Delta\chi}$, and denoting as $\mu_{jn}^{\Delta\chi}$ the $j$-th largest eigenvalue of $\bm\Gamma_n^{\Delta\chi}$, Assumptions \ref{ass:dynamic}(b) and \ref{ass:dynamic}(c) imply that, for $j=1,\ldots, q$,
\beq\label{eq:divevaldiff}
\underline K_j\le \lim\inf_{n\to\infty} n^{-1} \mu_{jn}^{\Delta\chi}\le\lim\sup_{n\to\infty} n^{-1} \mu_{jn}^{\Delta\chi}\le \overline K_j, 
\eeq
for some positive reals $\underline K_j$ and $\overline K_j$. Moreover,  letting the covariance matrix of $\Delta \bm \xi_n$ be $\bm\Gamma_n^{\Delta\xi}$, and denoting as $\mu_{jn}^{\Delta\xi}$ the $j$-th largest eigenvalue of $\bm\Gamma_n^{\Delta\xi}$, Assumption \ref{ass:modelNS}(c), implies that
\beq
\sup_{n\in\mathbb N} \mu_{1n}^{\Delta\xi}\le M_\xi, \label{eq:divevaldiff2}
\eeq
for some positive real $M_\xi$. From \eqref{eq:divevaldiff} and \eqref{eq:divevaldiff2}, and Assumption \ref{ass:modelNS}(e), by Weyl's inequality,  the $q$ largest eigenvalues of the covariance matrix of $\Delta \mbf x_{nt}$ diverge linearly in $n$, while all other $(n-q)$ eigenvalues stay bounded for all $n\in\mathbb N$. 

Moreover, it can be shown that the $q$ largest eigenvalues of the spectral density of $\Delta \mbf x_{nt}$ diverge with $n$ at all frequencies, but at zero-frequency, where, due to the presence of common trends, only $(q - d)$ eigenvalues diverge, all the others eigenvalues being bounded for all $n$ and all frequencies. Hence, by looking at the eigenvalues of the spectral density matrix of $\Delta \mbf x_{nt}$ we can determine $q$ and $d$ (see \citealp{hallinliska07}, and \citealp{BLL2}, respectively). Moreover, notice that when all factors are pervasive at all lags, i.e., in Assumption \ref{ass:dynamic}(b) we let $\mbox{rk}(\bm\Sigma_k)=q$ for all $k=0,\ldots,s$,  then \eqref{eq:divevaldiff} holds for all $j=1,\ldots,q(s+1)$.  Therefore, by looking at the eigenvalues of the covariance matrix of $\Delta \mbf x_{nt}$, we can also determine $s$ \citep{dagostinogiannone12}.

The model defined in \eqref{eq:NSDFM1}-\eqref{eq:NSDFM4} has $q$ latent states, given by the common factors $\bm f_t$, and additional latent states given by those idiosyncratic components that are autocorrelated as in \eqref{eq:NSDFM3}, and by the time-varying intercepts and trend slopes as in \eqref{eq:NSDFM5} and \eqref{eq:NSDFM4}. These additional latent states are such that they satisfy the following assumption. 

In other words, we are assuming that some, but not all, idiosyncratic components are $I(1)$, and that some, but not all, series have a time-varying intercept and/or a linear trend with time-varying slope. For simplicity, we are also assuming that stationary idiosyncratic components are serially uncorrelated. 

We then  make the following identifying assumptions.

\begin{ass} \label{ass:identNS}
Let $\mbf M_n^{\Delta \chi}$ be the $q\times q$ diagonal matrix with elements $\mu_{1n}^{\Delta \chi},\ldots,\mu_{qn}^{\Delta \chi}$, and let 
$\mbf V_n^{\Delta\chi}$ be the $n\times q$ matrix having as columns the corresponding normalized eigenvectors. Then: \begin{inparaenum}[(a)]
 \item $\Delta \bm f_t=(\mbf M_n^{\Delta\chi})^{-1/2}\mbf V_n^{\Delta\chi\prime} \Delta\bm\chi_{nt}$;
 \item the entries of $\mbf M_n^{\Delta\chi}$ are such that they satisfy \eqref{eq:divevaldiff} and $\overline K_{j+1}<\underline K_j$ for $j=1,\ldots, q-1$;
 \item the entries of $\mbf V_n^{\Delta\chi}$ are such that $[\mbf V_n^{\Delta\chi}]_{1j}>0$ for all $j=1,\ldots, q$.
\end{inparaenum}
\end{ass}
 
Parts (a) and (b) are standard in factor model literature for stationary processes and allow to identify the differenced factors up to a multiplication by a sign \citep[see, e.g.,][]{FGLR09,FLM13}. We identify the first difference of the factors with the $q$ normalized principal components of $\Delta\bm\chi_{nt}$ and this implies in Assumption \ref{ass:dynamic}(b) that $\bm\Gamma^{\Delta f}=\mbf I_q$. It can then be seen that the following must hold for the loadings
\begin{align}
\mbf V_n^{\Delta\chi\prime} \bm{\mathcal B}_{0n}= (\mbf M_n^{\Delta\chi})^{1/2},
\end{align}
therefore we can choose $\bm{\mathcal B}_{0n}=\mbf V_n^{\Delta\chi}(\mbf M_n^{\Delta\chi})^{1/2}$, and in Assumption \ref{ass:dynamic}(a) we have that $\bm\Sigma_0$ is diagonal with entries given by $\lim_{n\to\infty} (n^{-1}\mu_{jn}^{\Delta \chi})$, which as requested are finite and positive because of \eqref{eq:divevaldiff}. Part (c) is a way to fix the sign indeterminacy in the identification of the factors. Once $\Delta\bm f_t$ and $\bm{\mathcal B}_{0n}$ are identified, then the remaining loadings are obtained by projecting $\Delta\mbf x_{nt}$ onto the lagged factors.


The identifying restrictions in Assumption \ref{ass:identNS} are particularly useful for initializing the EM algorithm with the PC estimator (see the next section). However, it has to be stressed that this identification does not provide any economic meaning to the factors. In other words we are not interested here in giving any interpretation of the factors, but we are just interested in the common component, which is always identified.


%
%
\section{Estimation and asymptotic properties}\label{sec:estNS}
Throughout the rest of the section we assume to observe the $nT$-dimensional vector $\bm X_{nT}=(\mbf x_{n1}^\prime\cdots\mbf x_{nT}^\prime)^\prime$ satisfying \eqref{eq:NSDFM1}-\eqref{eq:NSDFM4}. In order to derive an estimator of the common component, we need to estimate the factors vector $\bm f_T=(\bm f_{1}^\prime\cdots \bm f_T^\prime)^\prime$ and the vector containing the true values of all parameters is 
 \[
 \bm\varphi_n=\l(\text{vec}(\bm{\mathcal B}_{0n}\cdots \bm{\mathcal B}_{sn})^\prime, \text{vech}(\bm\Gamma_n^e)^\prime, \rho_1,\ldots, \rho_{n_1} ,\text{vec}(\bm{\mathcal A}_1\cdots \bm{\mathcal A}_p)^\prime, \text{vech}(\bm\Gamma^u)^\prime,\sigma_{1\omega}^{2}\cdots\sigma^{2}_{n_a\omega},\sigma_{1\eta}^{2}\cdots\sigma^{2}_{n_b\eta}\r)^\prime, 
 \]
 where, without loss of generality, we assumed that $\mathcal I_1=\{1,\ldots, n_1\}$,  $\mathcal I_a=\{1,\ldots, n_a\}$, and  $\mathcal I_b=\{1,\ldots, n_b\}$.

In this Section, we provide asymptotic results when $n_1=0$, $n_a=0$, and $n_b=0$, thus assuming that all idiosyncratic component are stationary and that no time-varying term is present. At first sight this might seem as a strong requirement, but notice that in our framework introducing non-stationary idiosyncratic components and/or local levels and/or local linear trends implies just adding latent states. We discuss this extension in Section \ref{sec:idioI1}. Moreover, in \ref{app:prestNS}, we give all details of the EM algorithm together with explicit expressions for all estimators in the general case. 

Without loss of generality, we fix $s=1,$ and we fix the VAR order in \eqref{eq:NSDFM2} to $p=2$, thus $\bm{\mathcal A}(L)\equiv(\bm{\mathcal A}_1 L+\bm{\mathcal A}_2L^2)$, and, in this way the stationary component of $\bm f_t$ follows a non-trivial dynamics. For simplicity, we also assume that $\alpha_{i0}=0$ and $\beta_{i0}=0$. 

The EM algorithm is an iterative procedure, which starts with an initial value of the parameters $\wh{\bm\varphi}_n^{(0)}$, and at each iteration $k\ge 0$ produces estimates of the factors $\bm f_{t|T}^{(k)}$ (KS and E-step) and of the parameters $\wh{\bm\varphi}_n^{(k+1)}$ (M-step).  When the EM algorithm converges, say at iteration $k^*$, it gives the estimated common component $\wh{\chi}_{it}=\wh{\bm b}_{i0}^{(k^*+1)\prime}\bm f_{t|T}^{(k^*+1)}+\wh{\bm b}_{i1}^{(k^*+1)\prime}\bm f_{t-1|T}^{(k^*+1)}$. 

More in detail, the NS-DFM in \eqref{eq:NSDFM1}-\eqref{eq:NSDFM2} can be written as
\begin{align}
\mbf x_{nt}&=\l(\bm{\mathcal B}_{0n}\ \bm{\mathcal B}_{1n}\r)
\l(\ba{c}
\bm f_t\\
\bm f_{t-1}
\ea
\r)+\mbf e_{nt},\label{eq:SSNS001}\\
\l(\ba{c}
\bm f_t\\
\bm f_{t-1}
\ea
\r)&=\l(\ba{cc}
\bm{\mathcal A}_1&\bm{\mathcal A}_2\\
\mbf I_q&\mbf 0_{q\times q}
\ea
\r)
\l(\ba{c}
\bm f_{t-1}\\
\bm f_{t-2}
\ea
\r)
+\l(\ba{c}
\bm u_t\\
\mbf 0_q
\ea
\r),\label{eq:SSNS002}
\end{align} 
By defining  $\mbf F_t=(\bm f_t^\prime\; \bm f_{t-1}^\prime)^\prime$ and $\bm\lambda_i=(\bm{b}_{0i}^\prime\;\bm{b}_{1i}^\prime)^\prime$, we see that, for given values of the parameters $\wh{\bm\varphi}_n^{(k)}$, we can easily estimate the factors via the KS applied to the state-space form in \eqref{eq:SSNS001}-\eqref{eq:SSNS002}. The estimated states are then $\mbf F_{t|T}^{(k)}=\E_{\wh{\varphi}_n^{(k)}}[\mbf F_t|\bm X_{nT}]$, the first $q$-components of which give $\bm f_{t|T}^{(k)}=\E_{\wh{\varphi}_n^{(k)}}[\bm f_t|\bm X_{nT}]$. Then, using the output of the KS, we can compute the expected log-likelihood, which is maximized by the loadings estimator $\wh{\bm \lambda}_{i}^{(k+1)}\equiv(\wh{\bm b}_{i0}^{(k+1)\prime} \; \wh{\bm b}_{i1}^{(k+1)\prime})^\prime$, such that
\begin{align}
\wh{\bm \lambda}_{i}^{(k+1)}\!=
\l\{\sum_{t=1}^T\E_{\wh{\varphi}_n^{(k)}}\!\!\l[\l(\ba{ll}
\bm f_t\bm f_t^\prime&\bm f_t\bm f_{t-1}^\prime\\
\bm f_{t-1}\bm f_t^\prime&\bm f_{t-1}\bm f_{t-1}^\prime
\ea\r)\!\!\bigg\vert \bm X_{nT}
\r]\r\}^{\!\!-1}\!\!\!
\l\{\sum_{t=1}^T\E_{\wh{\varphi}_n^{(k)}}\!\!\l[\l(\ba{l}
\bm f_tx_{it}\\
\bm f_{t-1}x_{it}
\ea\r)\!\!\bigg\vert \bm X_{nT}
\r]\r\}.\label{eq:param1NS}
\end{align}

The initial value of the parameters $\wh{\bm\varphi}_n^{(0)}$ is determined as follows. For the loadings and the factors we use the approach proposed in \citet{BLL2}, which makes use of the $q$ leading PCs of the model in first differences. Two comments are worth making. 
First, it important to stress that initializing the model in first differences (including when determining $q$ and $s$) is crucial, since it allows us to use PCs without incurring in spurious effects due to the presence of idiosyncratic unit roots (\citealp{OW19}), or  linear trends (\citealp{ngCG}). Second, in light of the previous comment, this approach provides consistent estimates of the loadings, even in the case in which Assumption \ref{ass:I1} is satisfied with $n_1>0$ and $n_b>0$, but for constant intercepts and trend slopes \citep[see also][in the case of no linear trends]{baing04}. In particular, our initialization delivers estimates of $\alpha_{i0}$ and $\beta_{i0}$, which, together with a given small initial value of the variances $\wh{\sigma}_{i\omega}^{2(0)}$ and $\wh{\sigma}_{i\eta}^{2(0)}$,  can be used to update the slope state in \eqref{eq:NSDFM4}. Notice that the pre-estimators of those initial conditions do not need to be consistent for our results to hold.
The initialization is completed by estimating the parameters of  \eqref{eq:NSDFM2} from an unrestricted VAR fitted on the estimated factors. This is a valid procedure when estimating an autoregressive model for cointegrated data (see \citealp{simsstockwatson}). Consistency of the pre-estimators of the loadings and VAR coefficients is proved in \citet[Lemma 3 and Proposition 2]{BLL2} (see also \ref{app:consN}). 

Finally, notice also that we initialize the KF by setting the initial value of the covariance of the factors, ${\mbf P}_{0|0}$, to a very large value, as suggested by \citet[Section 3.3.4, page 121]{harvey90}. 

Consistency of the estimated common component follows.
\newpage
\begin{prop}\label{th:chiNS}
Under Assumptions \ref{ass:dynamic}, \ref{ass:modelNS}, \ref{ass:identNS}, and if $\mbox{rk}(\bm\Sigma_k)=q$ for all $k=0,\ldots,s$, and $n_1=0$, $n_a=0$, and $n_b=0$,  as $n,T\to\infty$, for any given $i=1,\ldots, n$ and $k=0,1$, $\min(\sqrt n,\sqrt T)\Vert \wh{\bm b}_{ki}-\bm b_{ki}\Vert = O_p(1)$, 
and, for any given $t=\bar t,\ldots, T$,
$\min(\sqrt n,\sqrt T)\Vert \wh{\bm f}_{t|T}-\bm f_{t}\Vert= O_p(1)$. Moreover, $\min(\sqrt n,\sqrt {T})\,\Vert \wh{\chi}_{it}-{\chi}_{it} \Vert = O_p(1)$,
for any given $i=1,\ldots,n$ and $t=\bar t,\ldots, T$, with $\bar t\ge 2$.
\end{prop}


The convergence rate depends on different ingredients. First, we show that the KS reaches a steady state within $\bar t$ periods, where $\bar t$ depends on the initial value $\mbf P_{0|0}$ and, as shown in Section \ref{sec:mc2}, $\bar t$ is typically very small.  Then, for the KS we show that,  given the true parameters, the factors are $\sqrt N$-consistent. Third, given the true factors, the loadings estimator are consistent, with convergence rate $T$ for the loadings of the $I(1)$ components of the factors and convergence rate $\sqrt T$ for the loadings of the stationary 
component of the factors. As a result, for any given $i$, the whole loadings vector is $\sqrt T$-consistent, unless $d=q$, in which case each all $q$ factors are random walks and then the loadings vector would be $T$-consistent. 

Under the assumption $n_1=0$, $n_a=0$, and $n_b=0$, our model is equivalent to the model studied in \citet{bai04}, who considers estimation by means of PCs in levels. In this respect, we notice that the rates in Proposition \ref{th:chiNS} are very similar to those in  \citet[Theorem 6 for the case $d<q$ and Theorem 4 for the case $d=q$]{bai04}. In other words, the QML estimator converges at the same rate than the estimator based on PC on the levels, which resembles the result in \citet{BLqml} for the stationary case. However, in the simulation study in Section \ref{sec:mc2} show that our estimator behave much better in finite samples.




%
%
\section{$I(1)$ idiosyncratic components, local levels, local linear trends}\label{sec:idioI1}

If some idiosyncratic components are non-stationary, we can no longer use the EM algorithm described in the previous section. Indeed, when the residuals of \eqref{eq:NSDFM1} are non-stationary, the M-step estimator of the loadings cannot be obtained by regressing $x_{it}$ is $I(1)$ onto  $\bm f_t$ and $\bm f_{t-1}$. However, the case in which some idiosyncratic components are $I(1)$ is the relevant one for large macroeconomic datasets, since otherwise all data would be cointegrated. This is, for example, shown by the empirical results in  \citet{OGAP}, where the methodology proposed by \citet{baing04} for testing for idiosyncratic unit roots is applied on a standard US macroeconomic dataset.

In this Section, we adapt the EM algorithm to model non-stationary idiosyncratic components as well as local levels and local linear trends. In particular, we borrow from the literature on nowcasting with stationary factor models which models autocorrelated idiosyncratic components by treating them as additional latent states (see, e.g., \citealp{banburamodugno14}; and \citealp{BGMR13}). 

Let us define $m=(n_1+n_a+n_b)$, as the number of additional latent states and recall that by Assumption \ref{ass:I1}, $n^{-1}m\to 0$, as $n\to\infty$. Define also the set $\mathcal I_{m}=\mathcal I_1\cup\mathcal I_a\cup\mathcal I_b$, and notice that $\#\mathcal I_m\le m$, since it is possible that a variable has both a non-stationary idiosyncratic component as well as, for example, a linear trend. Then for all $i\in\mathcal I_{m}$,  we replace the measurement equation \eqref{eq:NSDFM1} with
\begin{align}
x_{it}&=\alpha_{it}+\beta_{it}t + \bm b_i^\prime(L) \bm f_t + \xi_{it} + \nu_{it},\label{eq:NSDFM1bis}
\end{align}
such that Assumption \ref{ass:dynamic} still hold, and, moreover, letting $\bm\nu_{mt}=(\nu_{1t}\cdots\nu_{mt})^\prime$, for all $t\in\mathbb Z$, we assume $\bm\nu_{mt}\sim\mathcal N(\mbf 0_m,\phi\mbf I_m)$, with $\phi>0$, and $\E_{\varphi_n}[\bm\nu_{mt}\bm\nu_{mt-k}] = 0_{m\times m}$ for all $k\ne 0$. If $i\notin\mathcal I_m$, then \eqref{eq:NSDFM1} stays the same. Moreover, we leave the dynamics of the factors in \eqref{eq:NSDFM2} unchanged, while we change \eqref{eq:NSDFM3} to
\begin{align}
\xi_{it}&=\xi_{it-1} + e_{it},\;\text{ if }\; i\in \mathcal I_1,\;\text{ and }\; \xi_{it}= e_{it},\;\text{ if }\; i\notin \mathcal I_1,\label{eq:NSDFM3bis}
\end{align}
where Assumptions \ref{ass:modelNS}(b) and  \ref{ass:modelNS}(c) still hold, and $\E_{\varphi_n}[\nu_{it} e_{js}]=0$, for all $t,s\in\mathbb Z$, all $i\in\mathcal I_m$ and all $j=1,\ldots, n$.
Finally, according to \eqref{eq:NSDFM5} and \eqref{eq:NSDFM4}, we have the state equations
\begin{align}
&\alpha_{it}=\alpha_{it-1} + \omega_{it},\;\text{ if }\; i\in \mathcal I_a,\label{eq:NSDFM5bis}\\
&\beta_{it}=\beta_{it-1} + \eta_{it},\;\text{ if }\; i\in \mathcal I_b,\label{eq:NSDFM4bis}
\end{align}
such that $\E_{\varphi_n}[\nu_{it} \omega_{js}]=0$, and $\E_{\varphi_n}[\nu_{it} \eta_{js}]=0$, for all $t,s\in\mathbb Z$, all $i\in\mathcal I_m$, and  all $j\in\mathcal I_a$ or $j\in\mathcal I_b$.  

The model, which has as measurement equation either \eqref{eq:NSDFM1} or \eqref{eq:NSDFM1bis} if $i\in\mathcal I_m$, and which has as state equations  \eqref{eq:NSDFM2}, and, if needed, also equations \eqref{eq:NSDFM3bis}, \eqref{eq:NSDFM5bis} and \eqref{eq:NSDFM4bis}, has a compact state space form which is given in  \ref{app:prestNS}, together with the details on its estimation via the EM algorithm. In particular, letting $w_{it}=\alpha_{it}+\beta_{it}t+\xi_{it}$, for all $i\in\mathcal I_m$ we show that, at a given iteration $k\ge 0$ of the EM algorithm, the M-step gives the loadings estimators:
\begin{align}
\wh{\bm \lambda}_{i}^{(k+1)}\!=
\l\{\sum_{t=1}^T\E_{\wh{\varphi}_n^{(k)}}\!\!\l[\l(\ba{ll}
\bm f_t\bm f_t^\prime&\bm f_t\bm f_{t-1}^\prime\\
\bm f_{t-1}\bm f_t^\prime&\bm f_{t-1}\bm f_{t-1}^\prime
\ea\r)\!\!\bigg\vert \bm X_{nT}
\r]\r\}^{\!\!-1}\!\!\!
\l\{\sum_{t=1}^T\E_{\wh{\varphi}_n^{(k)}}\!\!\l[\l(\ba{l}
\bm f_t(x_{it}-w_{it})\\
\bm f_{t-1}(x_{it}-w_{it})
\ea\r)\!\!\bigg\vert \bm X_{nT},
\r]\r\}.\nn
\end{align}
where $\bm\lambda_i=(\bm{b}_{0i}^\prime\;\bm{b}_{1i}^\prime)^\prime$, while for $i\notin\mathcal I_m$ the loadings estimator is the same as in \eqref{eq:param1NS}. Formulas for all other estimators are given in \ref{app:prestNS}. In order to be able to compute $\wh{\bm \lambda}_{i}^{(k+1)}$, we have to estimate the $m$ additional latent states $w_{it}$ and therefore we also need modify the KS accordingly (see \ref{app:prestNS} for details). 

In \ref{app:misidioNS}, we provide an overview of the challenges involved by this task and we provide an informal derivation of the conditions necessary for consistent estimation, together with the related convergence rates.  Three main results emerge. First, the new latent states can be recovered only if they display also some degree of cross-sectional correlation, as if they were driven by some common factor which is weakly pervasive for the whole panel. The intuition is that, if the additional latent states are completely uncorrelated across the components of $\mbf x_{nt}$, then pooling many series does not help in recovering them, since their effect is always dominated by the factors. 

Second, when the previous condition is verified, then we can still achieve $\sqrt n$-consistency for the estimated factors (as in the proof of Proposition \ref{th:chiNS}), regardless of $m$, but provided that the variance of the measurement error $\nu_{it}$ in \eqref{eq:NSDFM1bis} is fixed in such a way that $\phi=o(n^{-1})$, that is, it is asymptotically negligible. Indeed, the presence of $\nu_{it}$ represents a mis-specification of the original model in \eqref{eq:NSDFM1}, which needs to be introduced only as a numerical device, since the KF is not be defined if $\phi=0$. The smaller is $\phi$, the smaller the effect of the mis-specification is, and, therefore, the estimation of the factors is unaffected by the additional states. 

As a consequence of this result, our estimator converges at a faster rate than those proposed by \citet{baing04} and \citet{BLL2}, which are based on PC analysis on the differenced data. This faster convergence rate comes from the fact that we distinguish \textit{a priori} between $I(1)$ and stationary idiosyncratic components. By contrast, due to differencing the estimator of \citet{baing04} and \citet{BLL2} essentially treat all idiosyncratic components as if they were $I(1)$. Of course, for the implementation of our estimator, it is crucial to be able to determine consistently which idiosyncratic component is $I(1)$---for example, using the test for idiosyncratic unit roots proposed by \citet{baing04}.

Third, to achieve consistency of the additional latent states a necessary condition is $mn^{-1}\to 0$. This reflects the obvious intuition that the more latent states we need to estimate, the worse the performance of our estimator is going to be. Moreover, $\sqrt n$-consistency for the new states can be obtained for any $m$, but only if we choose an even smaller value of $\phi$, namely $\phi=o((m\sqrt n)^{-1})$. 
 
We conclude with three remarks. First, the requirement that the new latent states display some degree of cross-sectional correlation is perfectly in line with Assumption \ref{ass:modelNS}(c) according to which the idiosyncratic components can be cross-correlated. Moreover, we can relax Assumptions \ref{ass:modelNS}(e) and \ref{ass:modelNS}(g) to allow for some correlation across the innovations $e_{it}$, $\omega_{it}$, and $\eta_{it}$ in \eqref{eq:NSDFM3bis}, \eqref{eq:NSDFM5bis} and \eqref{eq:NSDFM4bis}. Indeed, it is reasonable to assume that local linear trends are shared by real variables (e.g., GDP and GDI), or that local levels are more apt to capture time-varying mean of groups of variables belonging, for example, to the labor market. Nevertheless, as shown in the proof of Proposition \ref{th:chiNS}, the fact that we estimate the $I(1)$ idiosyncratic components without modeling the cross-correlation between their innovations, will add miss-specification to our model, but will not affect the consistency of our estimates. 

Second, as a far as estimation of the parameters given estimates of the states is concerned, we conjecture that nothing changes with respect to the results used in the proof of Proposition \ref{th:chiNS}, provided the states estimators are $\sqrt n$-consistent. Third, since the above are just  asymptotic arguments, the choice of $\phi$ is not straightforward. A common way to proceed consists in initializing $\phi$ to be very small for all $m$ additional states and then update its estimate at each iteration of the EM algorithm, thus adding $m$ additional parameters. This is the way we implement the EM algorithm in the next section (see also \ref{app:prestNS}).

\section{MonteCarlo results}\label{sec:mc2}
Throughout, we let $n\in\{75,100,200,300\}$, $T\in\{75,100,200,300\}$, $q\in\{2,4\}$, and $s\in\{0,1\}$, and we simulate data according to \eqref{eq:NSDFM1}, \eqref{eq:NSDFM2}, \eqref{eq:NSDFM3}, and \eqref{eq:NSDFM4} as follows.

First, the factor loadings are such that $[\bm{\mathcal B}_{kn}]_{ij} \sim N(1,1)$ for $k=0,\ldots, s$, and then if $s=1$, for all $j=1,\ldots q$, we take $n/2$ randomly selected elements of $[\bm{\mathcal B}_{1n}]_{\cdot j}$ and we set them to zero. Second, for the common factors we set the VAR order $p=2$, and to generate $\bm{\mathcal A}(L)$ we use the Smith-McMillan factorization according to which $\bm{\mathcal A}(L)=\mathbfcal{U}(L) \mathbfcal{M}(L) \mathbfcal{V}(L)$, where $\mathbfcal{M}(L)= \mbox{diag} \left( (1-L)\mbf I_{q-d}, \mbf I_d\right)$, $\mathbfcal{V}(L)=\mbf I_q$, and $\mathbfcal{U}(L)=(\mbf I_q-\mathbfcal{U}_1 L)$, where $\mathbfcal{U}_1=\mu\,\wt{\mathbfcal{U}}_1(\nu^{(1)}(\wt{\mathbfcal{U}}_1))^{-1}$, where the diagonal elements of $\wt{\mathbfcal{U}}_1$ are drawn from a uniform distribution on $[0.5,0.8]$, while the off-diagonal elements from a uniform distribution on $[0,0.3]$, and $\mu=0.5$. In this way, $\bm f_t$ follows a VAR(2) with $q-d$ unit roots, or, equivalently, a VECM(1), where the number cointegration relations is set to $d=1$. The common innovations are such that $\bm u_t\stackrel{iid}{\sim} \mathcal{N}(\mbf 0_q,\mbf I_q)$, or $\bm u_t\stackrel{iid}{\sim} t_4(\mbf 0_q,\mbf I_q)$.

Third, each idiosyncratic component follows an AR(2) with roots $\rho_{i1}$ and $\rho_{i2}$, such that $\rho_{i1}=1$ if $\xi_{it}\sim I(1)$, while $\rho_{i1}=0$ otherwise, and $\rho_{i2}$ is drawn from a uniform distribution on $[0.2,0.6]$. We randomly select $n_1$ idiosyncratic components to have a unit root, with $n_1\in\{0,25,50,75,100\}$, provided $n_1<n$. The innovations are such that $\mbf e_t\stackrel{iid}{\sim} \mathcal{N}(\mbf 0_n,\bm\Gamma^e_n)$, or $\mbf e_t\stackrel{iid}{\sim} t_4(\mbf 0_n,\bm\Gamma^e_n)$
 with $[\bm\Gamma^e_n]_{ij}=\tau^{|i-j|}$ if $\tau>0$, while, if $\tau=0$, $\bm\Gamma^e_n$ is diagonal with entries drawn from a uniform distribution on $[0.5,1.5]$. We set $\tau\in\{0,0.5\}$.

Fourth, we randomly select $n_b$ variables to have a non-zero linear trend, with $n_b\in\{0,25,50,75,100\}$, provided $n_b<n$. For those variables we draw $\beta_{i0}$ from a uniform distribution on ${[0.3,0.5]}$, but we set $\sigma_i^2=0$, thus considering only linear trends with constant slopes.

Last, we rescale the first differences of each common and idiosyncratic component in such a way that the share of variance of the $i$-th variable explained by the common component is $\theta(1+\theta)^{-1}$. We set $\theta=0.5$.


We consider $B=1000$ replications and we run the EM algorithm to estimate the NS-DFM by running the KS with $(q+n_1)$ latent states and estimating in the M-step only the diagonal terms of the idiosyncratic covariance matrix even when $\tau>0$. Similarly we do not add idiosyncratic states even when $\delta>0$. In other words, we always estimate a mis-specified model and, in this way, we are able to assess how robust our-estimators are with respect to mis-specifications. 

In Table \ref{tab:PttNS}, we report for different values of $n$ and for $t=1,\ldots, 10$, the trace of the one-step-ahead, KF, and KS MSEs when $q=2$, $s=1$, $T=100$, $\tau=0.5$, and $\delta=0.2$ (serially and cross-correlated idiosyncratic components). The MSEs are computed using the true simulated value of the parameters in order to verify numerically convergence to the steady-state. First, as $n$ grows, the one-step-ahead MSE reaches a steady state within maximum five time periods and $\mbox{tr}(\mbf P_{t|t-1})/q\simeq 1$. This is consistent with the fact that due to the presence of unit roots we inizialize the filter with a vary large value of $\mbf P_{0|0}$. Second, the KF and KS MSEs are very similar and both decrease to zero as $n$ grows and $\mbox{tr}(\mbf P_{t|t})n/q$ and $\mbox{tr}(\mbf P_{t|T})n/q$, computed when $t=10$, stabilize as $n$ grows thus showing that the rate of decrease is $n$.

\begin{table}[t!]
\setlength{\tabcolsep}{0\textwidth}
\caption{Simulation results NS-DFM}\label{tab:PttNS} \centering \smallskip

\textsc{Kalman filter and Kalman smoother MSEs} \smallskip

\scriptsize
\vskip .2cm
\begin{tabular}{L{.04\textwidth} L{.08\textwidth} | C{.11\textwidth}C{.11\textwidth}C{.11\textwidth}C{.1\textwidth}C{.11\textwidth}C{.11\textwidth}C{.11\textwidth}C{.11\textwidth}}
\multicolumn{10}{c}{Serially and cross-correlated idiosyncratic ($\tau=0.5$, $\delta=0.2$),  $n_1=0$, $n_b=0$, $q=2$, $s=1$}\\
\multicolumn{10}{c}{Gaussian innovations}\\[-4pt]
\\
\hline
\hline
&&\\[-4pt]
 	&$n$&	$5$	&	$10$	&	$25$	&	$50$	&	$75$	&	$100$	&	$200$	&	$300$	\\[4pt]
\hline
&&\\[-4pt]
\multicolumn{2}{l}{$\mbox{tr}(\mbf P_{0|0})/q$}	\vline 
	&	147.6161	&	154.1094	&	133.2495	&	152.7333	&	138.4666	&	109.5626	&	120.8897	&	134.7097	\\[4pt]
\hline
&&\\[-4pt]
&$t=1$	&	2.874610	&	1.601115	&	1.132925	&	1.076572	&	0.981511	&	1.020192	&	0.947337	&	0.980840	\\
&$t=2$	&	2.768656	&	1.543904	&	1.110419	&	1.064744	&	0.971928	&	1.014110	&	0.945185	&	0.979096	\\
\multirow{4}{*}{\rotatebox{90}{$\mbox{tr}(\mbf P_{t|t-1})/q$}}
&	$t=3$	&	2.748894	&	1.537389	&	1.109305	&	1.064269	&	0.970874	&	1.013479	&	0.945034	&	0.978928	\\
&	$t=4$	&	2.746597	&	1.536414	&	1.109239	&	1.064245	&	0.970735	&	1.013398	&	0.945022	&	0.978908	\\
&	$t=5$	&	2.746357	&	1.536248	&	1.109235	&	1.064244	&	0.970716	&	1.013387	&	0.945021	&	0.978906	\\
&	$t=6$	&	2.746329	&	1.536220	&	1.109235	&	1.064244	&	0.970714	&	1.013386	&	0.945020	&	0.978906	\\
&	$t=7$	&	2.746325	&	1.536215	&	1.109235	&	1.064244	&	0.970713	&	1.013385	&	0.945020	&	0.978906	\\
&	$t=8$	&	2.746324	&	1.536214	&	1.109235	&	1.064244	&	0.970713	&	1.013385	&	0.945020	&	0.978906	\\
&	$t=9$	&	2.746324	&	1.536214	&	1.109235	&	1.064244	&	0.970713	&	1.013385	&	0.945020	&	0.978906	\\
&	$t=10$	&	2.746324	&	1.536214	&	1.109235	&	1.064244	&	0.970713	&	1.013385	&	0.945020	&	0.978906	\\
[4pt]
\hline
&&\\[-4pt]
&	$t=1$	&	1.665780	&	0.548360	&	0.161240	&	0.107121	&	0.049408	&	0.036570	&	0.015982	&	0.011897	\\
&	$t=2$	&	1.386910	&	0.400762	&	0.116701	&	0.079362	&	0.037964	&	0.030533	&	0.011515	&	0.007687	\\
\multirow{4}{*}{\rotatebox{90}{$\mbox{tr}(\mbf P_{t|t})/q$}}
&	$t=3$	&	1.369813	&	0.393790	&	0.114352	&	0.078298	&	0.037113	&	0.030087	&	0.011295	&	0.007466	\\
&	$t=4$	&	1.366557	&	0.392627	&	0.114174	&	0.078221	&	0.037001	&	0.030029	&	0.011276	&	0.007446	\\
&	$t=5$	&	1.365967	&	0.392442	&	0.114163	&	0.078216	&	0.036985	&	0.030021	&	0.011274	&	0.007443	\\
&	$t=6$	&	1.365898	&	0.392411	&	0.114163	&	0.078216	&	0.036983	&	0.030020	&	0.011274	&	0.007443	\\
&	$t=7$	&	1.365890	&	0.392406	&	0.114163	&	0.078216	&	0.036983	&	0.030020	&	0.011274	&	0.007443	\\
&	$t=8$	&	1.365889	&	0.392405	&	0.114163	&	0.078216	&	0.036983	&	0.030020	&	0.011274	&	0.007443	\\
&	$t=9$	&	1.365889	&	0.392404	&	0.114163	&	0.078216	&	0.036983	&	0.030020	&	0.011274	&	0.007443	\\
&	$t=10$	&	1.365889	&	0.392404	&	0.114163	&	0.078216	&	0.036983	&	0.030020	&	0.011274	&	0.007443	\\
[4pt]
\hline
&&\\[-4pt]
\multicolumn{2}{l}{$\mbox{tr}(\mbf P_{10|10})n/q$}	\vline &	3.414722	&	1.962022	&	1.427032	&	1.955403	&	1.386863	&	1.501011	&	1.127376	&	1.116435	\\[4pt]
\hline
&&\\[-4pt]
&	$t=1$	&	0.938577	&	0.368873	&	0.110011	&	0.074161	&	0.028644	&	0.017577	&	0.010943	&	0.008560	\\
&	$t=2$	&	0.708279	&	0.253087	&	0.073988	&	0.051268	&	0.021393	&	0.014262	&	0.007372	&	0.005221	\\
\multirow{4}{*}{\rotatebox{90}{$\mbox{tr}(\mbf P_{t|T})/q$}}
&	$t=3$	&	0.697724	&	0.250786	&	0.072065	&	0.050450	&	0.020933	&	0.014096	&	0.007206	&	0.005061	\\
&	$t=4$	&	0.696501	&	0.250387	&	0.071913	&	0.050388	&	0.020872	&	0.014075	&	0.007192	&	0.005046	\\
&	$t=5$	&	0.696180	&	0.250328	&	0.071903	&	0.050384	&	0.020864	&	0.014072	&	0.007190	&	0.005045	\\
&	$t=6$	&	0.696142	&	0.250318	&	0.071903	&	0.050384	&	0.020863	&	0.014072	&	0.007190	&	0.005045	\\
&	$t=7$	&	0.696138	&	0.250317	&	0.071903	&	0.050384	&	0.020863	&	0.014072	&	0.007190	&	0.005045	\\
&	$t=8$	&	0.696137	&	0.250316	&	0.071903	&	0.050384	&	0.020863	&	0.014072	&	0.007190	&	0.005045	\\
&	$t=9$	&	0.696137	&	0.250316	&	0.071903	&	0.050384	&	0.020863	&	0.014072	&	0.007190	&	0.005045	\\
&	$t=10$	&	0.696137	&	0.250316	&	0.071903	&	0.050384	&	0.020863	&	0.014072	&	0.007190	&	0.005045	\\
[4pt]
\hline
&&\\[-4pt]											
\multicolumn{2}{l}{$\mbox{tr}(\mbf P_{10|T})n/q$}	\vline &	1.740343	&	1.251582	&	0.898783	&	1.259595	&	0.782350	&	0.703592	&	0.719033	&	0.756698	\\[4pt]
\hline
\hline

\end{tabular}
\end{table}

In Table \ref{tab:mc1NS} and in Table \ref{tab:mc1NSt}, we report the relative MSE of our estimator over the MSE of the common component estimators obtained by PC as in \citet{bai04}, and by PC in first differences as in \citet{baing04} and \citet{BLL2}. Overall our estimator outperforms the others with the exception of the latter, which is show to perform better when $n_1$ becomes very large and about the same order of magnitude as $n$. This reflects the additional computational burden of our estimator which requires increasing the number of latent states when the idiosyncratic components are non-stationary and therefore we must include their dynamics in the model.

\begin{table}[ht!]
\setlength{\tabcolsep}{0\textwidth}
\caption{Simulation results - Common components}\label{tab:mc1NS}
\centering
\small \textsc{Relative Mean Squared Errors} \smallskip

\footnotesize
\begin{tabular}{C{.085\textwidth}C{.085\textwidth}C{.085\textwidth}C{.085\textwidth} | C{.11\textwidth} C{.11\textwidth} C{.11\textwidth} | C{.11\textwidth} C{.11\textwidth} C{.11\textwidth} }
\multicolumn{10}{c}{Serially and cross correlated idiosyncratic components ($\tau=0.5$, $\delta =0.2$). Gaussian innovations}\\
\hline
\hline
&&&&\multicolumn{3}{c|}{$q=2$, $s=0$}&\multicolumn{3}{c}{$q=2$, $s=1$}\\\hline
$n$ & $T$ & $n_1$& $n_b$& \tiny{B}& \tiny{BN}& \tiny{BLL}& \tiny{B}& \tiny{BN}& \tiny{BLL}\\
\hline
75	&	75	&	0	&	0	&	0.58	&	0.00	&	0.28	&	0.54	&	0.01	&	0.53	\\
100	&	100	&	0	&	0	&	0.54	&	0.00	&	0.22	&	0.54	&	0.00	&	0.49	\\
200	&	200	&	0	&	0	&	0.45	&	0.00	&	0.12	&	0.59	&	0.00	&	0.39	\\
300	&	300	&	0	&	0	&	0.40	&	0.00	&	0.08	&	0.63	&	0.00	&	0.32	\\
\hline
75	&	75	&	25	&	25	&	0.01	&	0.02	&	0.44	&	0.04	&	0.04	&	0.83	\\
100	&	100	&	25	&	25	&	0.01	&	0.01	&	0.47	&	0.02	&	0.02	&	0.68	\\
200	&	200	&	25	&	25	&	0.00	&	0.00	&	0.53	&	0.01	&	0.01	&	0.66	\\
300	&	300	&	25	&	25	&	0.00	&	0.00	&	0.77	&	0.00	&	0.00	&	0.73	\\
\hline
75	&	75	&	50	&	50	&	0.05	&	0.21	&	1.55	&	0.11	&	0.23	&	1.47	\\
100	&	100	&	50	&	50	&	0.02	&	0.12	&	1.45	&	0.06	&	0.12	&	1.53	\\
200	&	200	&	50	&	50	&	0.00	&	0.02	&	0.82	&	0.01	&	0.02	&	0.75	\\
300	&	300	&	50	&	50	&	0.00	&	0.01	&	0.93	&	0.00	&	0.01	&	0.80	\\
\hline
100	&	100	&	75	&	75	&	0.03	&	0.23	&	1.44	&	0.08	&	0.23	&	1.48	\\
200	&	200	&	75	&	75	&	0.00	&	0.03	&	0.80	&	0.02	&	0.06	&	1.52	\\
300	&	300	&	75	&	75	&	0.00	&	0.02	&	1.11	&	0.01	&	0.02	&	0.92	\\
\hline
200	&	200	&	100	&	100	&	0.01	&	0.12	&	1.87	&	0.04	&	0.15	&	2.08	\\
300	&	300	&	100	&	100	&	0.00	&	0.03	&	1.08	&	0.01	&	0.04	&	1.45	\\
\hline
\hline
\end{tabular}

\begin{tabular}{p{\textwidth}}\tiny
This table reports relative MSEs of the QML estimator proposed in this paper over the MSE of the common component estimators obtained by PC as in \citet{bai04} (B), and by PC in first differences as in \citet{baing04} (BN) and \citet{BLL2} (BLL).
\end{tabular}
\end{table}

\begin{table}[t!]
\setlength{\tabcolsep}{0\textwidth}
\caption{Simulation results NS-DFM - Common components}\label{tab:mc1NSt}
\centering
\small \textsc{Relative Mean Squared Errors} \smallskip

\footnotesize
\begin{tabular}{C{.085\textwidth}C{.085\textwidth}C{.085\textwidth}C{.085\textwidth} | C{.11\textwidth} C{.11\textwidth} C{.11\textwidth} | C{.11\textwidth} C{.11\textwidth} C{.11\textwidth} }
\multicolumn{10}{c}{Serially and cross correlated idiosyncratic components ($\tau=0.5$, $\delta =0.2$). Student $t_4$ innovations}\\
\hline
\hline
&&&&\multicolumn{3}{c|}{$q=2$, $s=0$}&\multicolumn{3}{c}{$q=2$, $s=1$}\\
$n$ & $T$ & $n_1$& $n_b$&Rel-MSE&Rel-MSE&Rel-MSE&Rel-MSE&Rel-MSE&Rel-MSE\\
&&&&\tiny B&\tiny BN&\tiny BLL&\tiny B&\tiny BN&\tiny BLL\\
\hline
75	&	75	&	0	&	0	&	0.58	&	0.00	&	0.30	&	0.58	&	0.01	&	0.57	\\
100	&	100	&	0	&	0	&	0.55	&	0.00	&	0.25	&	0.56	&	0.00	&	0.53	\\
200	&	200	&	0	&	0	&	0.45	&	0.00	&	0.12	&	0.60	&	0.00	&	0.40	\\
300	&	300	&	0	&	0	&	0.41	&	0.00	&	0.09	&	0.67	&	0.00	&	0.35	\\
\hline
75	&	75	&	25	&	25	&	0.01	&	0.02	&	0.43	&	0.06	&	0.04	&	0.83	\\
100	&	100	&	25	&	25	&	0.01	&	0.01	&	0.39	&	0.03	&	0.02	&	0.70	\\
200	&	200	&	25	&	25	&	0.00	&	0.00	&	0.40	&	0.01	&	0.00	&	0.64	\\
300	&	300	&	25	&	25	&	0.00	&	0.00	&	0.56	&	0.01	&	0.00	&	0.68	\\
\hline
75	&	75	&	50	&	50	&	0.06	&	0.17	&	1.51	&	0.14	&	0.17	&	1.45	\\
100	&	100	&	50	&	50	&	0.04	&	0.11	&	1.57	&	0.09	&	0.10	&	1.46	\\
200	&	200	&	50	&	50	&	0.00	&	0.01	&	0.57	&	0.01	&	0.01	&	0.69	\\
300	&	300	&	50	&	50	&	0.00	&	0.01	&	0.63	&	0.01	&	0.01	&	0.71	\\
\hline																			
100	&	100	&	75	&	75	&	0.04	&	0.18	&	1.36	&	0.12	&	0.21	&	1.48	\\
200	&	200	&	75	&	75	&	0.01	&	0.02	&	0.65	&	0.03	&	0.05	&	1.41	\\
300	&	300	&	75	&	75	&	0.00	&	0.01	&	0.71	&	0.01	&	0.01	&	0.81	\\
\hline
200	&	200	&	100	&	100	&	0.02	&	0.10	&	1.68	&	0.06	&	0.13	&	1.99	\\
300	&	300	&	100	&	100	&	0.00	&	0.02	&	0.72	&	0.02	&	0.04	&	1.39	\\
\hline
\hline
\end{tabular}

\begin{tabular}{p{\textwidth}}\tiny
This table reports relative MSEs of the QML estimator proposed in this paper over the MSE of the common component estimators obtained by PC as in \citet{bai04} (B), and by PC in first differences as in \citet{baing04} (BN) and \citet{BLL2} (BLL).
\end{tabular}
\end{table}

Some clarifications on the competing methods considered are necessary in order to interpret the results in Table \ref{tab:mc1NS} and in Table \ref{tab:mc1NSt} (we refer to the original papers for details). First, notice that all alternative approaches considered here do not allow for dynamic loadings, so here they are implemented by computing the first $q(s+1)$ PCs. 

Second, despite the common practice in the literature, \citet{bai04} did not propose its approach for factor model estimation, but rather to estimate common trends, and it is based on the crucial assumption of all idiosyncratic components being stationary. Indeed, we see from Table \ref{tab:mc1NS} that when $n_1>0$ this approach fails completely. 

Third, the \citet{baing04} approach delivers estimates of the common component which are obtained 
\begin{inparaenum}[($i$)]
	\item by  detrending the data by estimating the slope of the trend with the mean of the data in first difference; then
	\item by estimating the factors in first differences; and, finally,
	\item by cumulating the differenced estimator to obtain an estimate of the levels. 
\end{inparaenum} 
As such, this estimator is always subject to a location shift---it can be shown to converge to a Brownian bridge. Notice that this approach was introduced to test for the presence of unit roots rather than for factor model estimation, and, while the test is unaffected by location shifts, the use of the cumulated estimator for other scopes is not justified in general. As we see from Table \ref{tab:mc1NS}, this approach fails to consistently reconstruct the common component in all cases considered. 

Fourth, the approach in \citet{BLL2} is based on the same ideas of \citet{baing04}, but it takes care of the above mentioned issues related to detrending and cumulation, and, therefore, it is a valid alternative.

%
%
\section{Concluding remarks}\label{sec:conclusion}

This paper considers estimation of large non-stationary approximate dynamic factor models by means of the Expectation Maximization algorithm, implemented jointly with the Kalman smoother. In our model the factors are a cointegrated vector process, thus containing both common $I(1)$ trends and stationary (cyclical) components.  We show that, as the cross-sectional dimension $n$ and the sample size $T$ diverge to infinity, the common factors, the factor loadings, and the common component estimated are $\min(\sqrt n,\sqrt T)$-consistent at each $i$ and $t$.

Furthermore, we show that the model can be extended to account for the possible presence of idiosyncratic trends, as well as the presence of secular (linear) trends, which can have either a constant slope (deterministic linear trends) or a time-varying slope (local linear trends). Consistent estimation of this case is also considered. 

Finally, the results in this paper provides the theoretical background for the application considered in \citet{OGAP}, where the NS-DFM  is used to estimate the output gap in the US.

\singlespacing
{\small{
\setlength{\bibsep}{.2cm}
\bibliographystyle{chicago}
\bibliography{BL_biblio}
}}
\setcounter{section}{0}%
\setcounter{subsection}{0}
\setcounter{equation}{0}
\setcounter{table}{0}
\setcounter{figure}{0}
\setcounter{footnote}{0}
\gdef\thesection{Appendix \Alph{section}}
\gdef\thesubsection{\Alph{section}.\arabic{subsection}}
\gdef\thefigure{\Alph{section}\arabic{figure}}
\gdef\theequation{\Alph{section}\arabic{equation}}
\gdef\varphible{\Alph{section}\arabic{table}}
\gdef\thefootnote{\Alph{section}\arabic{footnote}}

%
%
\clearpage
\small
\section{Estimation in practice}\label{app:prestNS}
Throughout, for simplicity, and without loss of generality, we let $p=2$ in the VAR for the factors \eqref{eq:NSDFM2}.

\subsection{State space representation}

Define $\mbf F_t=(\bm f_t^\prime\cdots \bm f_{t-s}^\prime)^\prime$ be the $r$-dimensional vector of factors and $\bm\Lambda_n = (\bm {\mathcal B}_{0n}\cdots \bm {\mathcal B}_{sn})$ be the $n\times r$ matrix containing the factor loadings at all $s$ lags. Define also 
\[
\mbf A = \l(\ba{cc}
\bm {\mathcal A}_{1}& \bm {\mathcal A}_{2}\\
\mbf I_q &\mbf 0_{q\times 0}
\ea\r), \qquad 
\mbf H =\l(\ba{c}
\mbf I_q\\
\mbf 0_{q\times q}
\ea
\r), \qquad 
\mbf R_n=\l(\ba{cccc}
\rho_1&\cdots&\cdots& 0\\
0&\rho_2&\cdots& 0\\
\vdots&\vdots&\ddots&\vdots\\
0&\cdots&\cdots& \rho_n
\ea
\r)
\] 
Define also $\bm\alpha_{nt}=(\alpha_{1t}\cdots \alpha_{nt})^\prime$ and $\bm\beta_{nt}=(\beta_{1t}\cdots \beta_{nt})^\prime$.
Then, using the process $\bm \nu_{nt}=(\nu_{1t}\cdots \nu_{nt})^\prime$ (see Section \ref{sec:idioI1}), we have the state space form for any $t=1,\ldots, T$:
\begin{align}
&\mbf x_{nt}=\l(\bm\Lambda_n\; \bm{\mathcal S}_{1n}\; \bm{\mathcal S}_{an} \; \bm{\mathcal S}_{bn}\r)
\l(\ba{c}
\mbf F_t\\
\bm\xi_{nt}\\
\bm\alpha_{nt}\\
\bm\beta_{nt}t
\ea
\r)+\bm \nu_{nt},\label{practice_obs}\\
&\l(\ba{c}
\mbf F_t\\
\bm\xi_{nt}\\
\bm\alpha_{nt}\\
\bm\beta_{nt}t
\ea
\r)=
\l(\ba{cccc}
\mbf A&\mbf 0_{r\times n}&\mbf 0_{r\times n}&\mbf 0_{r\times n}\\
\mbf 0_{n\times r}&\mbf R_n&\mbf 0_{n\times n}&\mbf 0_{n\times n}\\
\mbf 0_{n\times r}&\mbf 0_{n\times n}&\mbf I_n&\mbf 0_{n\times n}\\
\mbf 0_{n\times r}&\mbf 0_{n\times n}&\mbf 0_{n\times n}&\mbf I_n\\
\ea
\r)
\l(\ba{c}
\mbf F_{t-1}\\
\bm\xi_{nt-1}\\
\bm\alpha_{nt-1}\\
\bm\beta_{nt-1}
\ea
\r)+
\l(\ba{cccc}
\mbf H&\mbf 0_{r\times n}&\mbf 0_{r\times n}&\mbf 0_{r\times n}\\
\mbf 0_{n\times q}&\mbf I_n&\mbf 0_{n\times n}&\mbf 0_{n\times n}\\
\mbf 0_{n\times q}&\mbf 0_{n\times n}&\mbf I_n&\mbf 0_{n\times n}\\
\mbf 0_{n\times q}&\mbf 0_{n\times n}&\mbf 0_{n\times n}&\mbf I_n\\
\ea
\r)
\l(\ba{c}
\bm u_{t}\\
\mbf e_{nt}\\
\bm\omega_{nt}\\
\bm\eta_{nt}
\ea
\r),\nn
\end{align}
where the innovations are such that
\begin{align}
\l(\ba{c}
\bm\nu_{nt}\\
\bm u_{t}\\
\mbf e_{nt}\\
\bm\omega_{nt}\\
\bm\eta_{nt}
\ea
\r)\sim \mathcal N\l(
\l(\ba{c}
\mbf 0_n\\
\mbf 0_q\\
\mbf 0_n\\
\mbf 0_n\\
\mbf 0_n
\ea
\r),\l(\ba{ccccc}
\bm\Gamma_n^\nu&\mbf 0_{n\times q}&\mbf 0_{n\times n}&\mbf 0_{n\times n}&\mbf 0_{n\times n}\\
\mbf 0_{n\times n}&\bm\Gamma^u&\mbf 0_{n\times n}&\mbf 0_{n\times n}&\mbf 0_{n\times n}\\
\mbf 0_{n\times n}&\mbf 0_{n\times q}&\bm\Gamma_n^e&\mbf 0_{n\times n}&\mbf 0_{n\times n}\\
\mbf 0_{n\times n}&\mbf 0_{n\times q}&\mbf 0_{n\times n}&\bm\Gamma_n^\omega&\mbf 0_{n\times n}\\
\mbf 0_{n\times n}&\mbf 0_{n\times q}&\mbf 0_{n\times n}&\mbf 0_{n\times n}&\bm\Gamma_n^{\eta}\\
\ea
\r)
\r).\nn
\end{align}
where $\bm{\mathcal S}_{1n}$, $\bm{\mathcal S}_{an}$, $\bm{\mathcal S}_{bn}$ are $n\times n$ diagonal matrices with entries $\{0,1\}$, $\bm\Gamma_n^\nu$, $\bm\Gamma_n^\omega$, and $\bm\Gamma_n^\eta$ are $n\times n$ diagonal matrices with entries $\sigma_{i\nu}^2$,
$\sigma_{i\omega}^2$, and $\sigma_{i\eta}^2$, respectively,  $\bm\Gamma^u$ satisfies Assumption \ref{ass:modelNS}(a), and 
 $\bm\Gamma_n^{e}$ satisfies Assumptions \ref{ass:modelNS}(b) and \ref{ass:modelNS}(c).
Specifically, letting $\mathcal I_m=\mathcal I_1\cup \mathcal I_a\cup \mathcal I_b$, the following constraints apply:

\noindent 
\begin{center}
\begin{tabular}{llll}
\hline
\hline
&\\[-5pt]
$i\in\mathcal I_1$		&$[\bm{\mathcal S}_{1n}]_{ii}=1$  	&$\rho_i =1$				& $\sigma_{i\nu}^2>0$	\\
$i\notin\mathcal I_1$		&$[\bm{\mathcal S}_{1n}]_{ii}=0$ 	& $\rho_i =0$									\\
$i\in\mathcal I_a$ 		&$[\bm{\mathcal S}_{an}]_{ii}=1$ 	&$\sigma_{i\omega}^2>0$		& $\sigma_{i\nu}^2>0$ 	\\
$i\notin\mathcal I_a$		&$[\bm{\mathcal S}_{an}]_{ii}=0$  	&$\sigma_{i\omega}^2=0$ 						\\
$i\in\mathcal I_b$		&$[\bm{\mathcal S}_{bn}]_{ii}=1$  	&$\sigma_{i\eta}^2>0$		& $\sigma_{i\nu}^2>0$ 	\\
$i\notin\mathcal I_b$		&$[\bm{\mathcal S}_{bn}]_{ii}=0$  	&$\sigma_{i\eta}^2=0$ 							\\
$i\notin \mathcal I_m$	& 							&						&$\sigma_{i\nu}^2=[\bm\Gamma_n^e]_{ii}$\\[4pt]
\hline
\hline
\end{tabular}
\end{center}
In a more compact form equation the state space model \eqref{practice_obs} can be rewritten as
\begin{align}
\mbf x_{nt}&= \bm \Upsilon_n \bm s_{nt}+\bm \nu_{nt},\label{practice_obs2}.\\
\bm s_{nt} &= \bm\Theta_n \bm s_{nt-1}+\bm\zeta_{nt},\nn
\end{align}
with obvious definitions of $\bm \Upsilon_n$,  $\bm s_{nt}$, $ \bm\Theta_n$, and $\bm\zeta_{nt}$. This model  is equivalent to \eqref{eq:NSDFM1}-\eqref{eq:NSDFM4} up to the error term $\bm\nu_{nt}$ which is needed to run the KS and notice that for those series such that $i\notin \mathcal I_1\cup \mathcal I_a\cup \mathcal I_b$, then we are setting $\nu_{it}=0$ and therefore $\xi_{it}=e_{it}$ is the measurement equation error. In other words $\bm\Gamma_n^\nu$ is always positive definite. As explained in Section \ref{sec:idioI1}, this term is controlled by means of its variance $\phi$ and the smaller this is the better rate of convergence.

\subsection{Initialization}
Hereafter, for simplicity and without loss of generality, we let $s=1$, so that $r=q(s+1)=2q$, $\mbf F_t=(\bm f_t^\prime\; \bm f_{t-1}^\prime)^\prime$ and $\bm\Lambda_n = (\bm{\mathcal B}_{0n}\; \bm {\mathcal B}_{1n})$.

The pre-estimators are defined as follows. 
Let $\wh{\bm\Gamma}_n^{\Delta x}$ be the sample covariance matrix of the differenced data $\Delta \mbf x_{nt}$ and denote as $\wh{\mbf M}_n^{\Delta x}$ the diagonal matrix with entries the $q$-largest eigenvalues of $\wh{\bm\Gamma}_n^{\Delta x}$, and as $\wh{\mbf V}_n^{\Delta x}$ the $n\times q$ matrix of the corresponding normalized eigenvectors. We have the following pre-estimator of the loadings:
\begin{align}
&\wh{\bm{\mathcal B}}_{0n}^{(0)}= \wh{\mbf V}_n^{\Delta x}(\wh{\mbf M}_n^{\Delta x})^{1/2}, \nn
\end{align}
For all $i\in\mathcal I_a\cup\mathcal I_b$, let $\check{\alpha}_i$ and $\check{\beta}_i$ be the estimated parameter obtained by least squares of  $x_{it}$ onto a constant and a time trend, and let  $\check{x}_{it}=x_{it}-\check{\alpha}_i-\check{\beta}_it$. If $i\notin\mathcal I_a\cup\mathcal I_b$ define $\check {\alpha}_i=0$ and $\check{\beta}_i=0$. Then define: $\check{\mbf x}_{nt}=(\check x_{1t}\cdots \check x_{nt})^\prime$. The pre-estimator of the factors is given by
\beq
\wt{\bm f}_t = (\wh{\mbf M}_n^{\Delta x})^{-1}\wh{\bm{\mathcal B}}_{0n}^{(0)\prime}\check{\mbf x}_{nt}.\nn
\eeq
Moreover, we define 
\beq
\wh{\bm{\mathcal B}}_{1n}^{(0)} =\l(\sum_{t=2}^T (\Delta\mbf x_{nt}- \wh{\bm{\mathcal B}}_{0n}^{(0)}\Delta\wt{\bm f}_{t})\Delta\wt{\bm f}_{t-1}^\prime \r) \l(\sum_{t=2}^T \Delta\wt{\bm f}_{t-1} \Delta\wt{\bm f}_{t-1}^\prime \r)^{-1}.\nn
\eeq
Then, letting $\wt{\mbf F}_{t}=(\wt{\bm f}_t^\prime\,\wt{\bm f}_{t-1}^\prime)^\prime$, we define
\begin{align}
&\wh{\mbf A}^{(0)}=  \l(\sum_{t=3}^T\l(\ba{cc}
\wt{\bm f}_t\wt{\bm f}_{t-1}^\prime&\wt{\bm f}_t\wt{\bm f}_{t-2}^\prime\\
\wt{\bm f}_{t-1}\wt{\bm f}_{t-1}^\prime&\wt{\bm f}_{t-1}\wt{\bm f}_{t-2}^\prime
\ea
\r) \r)
\l(\sum_{t=3}^T 
\l(\ba{cc}
\wt{\bm f}_{t-1}\wt{\bm f}_{t-1}^\prime&\wt{\bm f}_{t-1}\wt{\bm f}_{t-2}^\prime\\
\wt{\bm f}_{t-2}\wt{\bm f}_{t-1}^\prime&\wt{\bm f}_{t-2}\wt{\bm f}_{t-2}^\prime
\ea
\r)
\r)^{-1}.\nn
\end{align}
and $\wh{\bm {\mathcal A}}_1^{(0)}$ is top left $q\times q$ block of $\wh{\mbf A}^{(0)}$, while $\wh{\bm {\mathcal A}}_2^{(0)}$ is top right $q\times q$ block of $\wh{\mbf A}^{(0)}$. Also, we define
\[
\wh{\bm\Gamma}^{u(0)}=\frac 1T\sum_{t=3}^T(\wt{\bm f}_t-\wh{\bm {\mathcal A}}_1^{(0)}\wt{\bm f}_{t-1}-\wh{\bm {\mathcal A}}_2^{(0)}\wt{\bm f}_{t-2})(\wt{\bm f}_t-\wh{\bm {\mathcal A}}_1^{(0)}\wt{\bm f}_{t-1}-\wh{\bm {\mathcal A}}_2^{(0)}\wt{\bm f}_{t-2})^\prime.
\]
Moreover, letting $\wh{\bm b}_{0i}^{(0)\prime}$ and $\wh{\bm b}_{1i}^{(0)\prime}$ be the $i$-th row of $\wh{\bm{\mathcal B}}_{0n}^{(0)}$ and of $\wh{\bm{\mathcal B}}_{1n}^{(0)}$, respectively, 
\begin{align}
[\wh{\bm\Gamma}_n^{e(0)}]_{ii}&=
\frac 1 {T}\sum_{t=2}^T
 \l(\Delta x_{it} - \wh{\bm b}_{0i}^{(0)\prime} \Delta\wt{\bm f}_{t}-\wh{\bm b}_{1i}^{(0)\prime}\Delta\wt{\bm f}_{t-1} \r)^2,\qquad i\in\mathcal I_1,\nn\\
[\wh{\bm\Gamma}_n^{e(0)}]_{ii}&=
\frac 1 {2T}\sum_{t=2}^T
 \l(\Delta x_{it} - \wh{\bm b}_{0i}^{(0)\prime} \Delta\wt{\bm f}_{t}-\wh{\bm b}_{1i}^{(0)\prime}\Delta\wt{\bm f}_{t-1}  \r)^2,\qquad i\notin\mathcal I_1,\nn
 \end{align}
 while $[\wh{\bm\Gamma}_n^{e(0)}]_{ij}=0$ if $i\ne j$.\footnote{Alternatively, when $ i\notin\mathcal I_1$, we can set
$[\wh{\bm\Gamma}_n^{e(0)}]_{ii}=
T^{-1}\sum_{t=2}^T
 ( x_{it} - \wh{\bm b}_{0i}^{(0)\prime} \wt{\bm f}_{t}-\wh{\bm b}_{1i}^{(0)\prime}\wt{\bm f}_{t-1}  )^2$. } 

Finally, if $i\in\mathcal I_a$ we define $\wh{\alpha}_{i0}^{(0)}=\check{\alpha}_i$ and if $i\in\mathcal I_b$ we define $\wh{\beta}_{i0}^{(0)}=\check{\beta}_i$, while $\wh{\sigma}^{2(0)}_{i\omega}=10^{-2}$ and $\wh{\sigma}^{2(0)}_{i\eta}=10^{-2}$, while if $i\in\mathcal I_m$, we fix $\wh{\sigma}_{i\nu}^{2(0)}=10^{-5}$.

All the above quantities are collected into the vector of initial estimates of the parameters $\wh{\bm\varphi}_n^{(0)}$.

\subsection{E-step}
To compute the expected log-likelihood of the model we run the KF-KS for the model in \eqref{practice_obs} or \eqref{practice_obs2}. The iterations of the KF-KS are standard and not reported. We just notice that, at iteration $k=0$ of the EM algorithm, the KF is inizialized as follows: we set $\bm f_{0|0}=\wt{\bm f}_0$ and, letting $\check{\mbf A}^{(0)}=0.99 {\wh{\mbf A}^{(0)}}({\Vert\wh{\mbf A}^{(0)}\Vert})^{-1}$, we set
\beq
{\mbf P}^{(0)}_{0|0}=\text{vec}^{-1}\l((\mbf I_{r^2}-\check{\mbf A}^{(0)}\otimes \check{\mbf A}^{(0)})^{-1}\text{vec}(\wh{\bm\Gamma}^{u(0)})\r).
\eeq 
Then, at each iteration $k\ge 0$ the EM algorithm produces estimates 
of all states are computed using the parameters $\wh{\bm\varphi}_n^{(k)}$ via KS. We obtain a vector $\bm s_{nt|T}^{(k)}=(\bm f_{t|T}^{(k)\prime}\, \bm\xi_{nt|T}^{(k)\prime}\,\bm\alpha_{nt|T}^{(k)\prime}\, \bm\beta_{nt|T}^{(k)\prime})^\prime$ with $(q+n_1+n_a+n_b)$ elements, such that 
\noindent 
\begin{center}
\begin{tabular}{llll}
\hline
\hline
&\\[-5pt]
$\bm f_{t|T}^{(k)}$		& 	$f_{j,t|T}^{(k)}$ 		& $j=1,\ldots ,q$\\
$\bm f_{t-1|T}^{(k)}$ 		& 	$f_{j,t-1|T}^{(k)}$ 	& $j=1,\ldots ,q$\\
$\bm\xi_{nt|T}^{(k)}$		& 	$\xi_{it|T}^{(k)}$ 	& if $i\in\mathcal I_1$\\
					& 	0				& if $i\notin\mathcal I_1$ \\
$\bm\alpha_{nt|T}^{(k)}$	& 	$\alpha_{it|T}^{(k)}$ 	& if $i\in\mathcal I_a$\\
					& 	0				& if $i\notin\mathcal I_a$ \\
$\bm\beta_{nt|T}^{(k)}$	& 	$\beta_{it|T}^{(k)}$ 	& if $i\in\mathcal I_b$\\
					& 	0				& if $i\notin\mathcal I_b$ \\
\hline
\hline
\end{tabular}
\end{center}
Finally, let $\bm w_{nt|T}^{(k)}=\bm\alpha_{nt|T}^{(k)}+\bm\beta_{nt|T}^{(k)}t+\bm \xi_{nt|T}^{(k)}$, which is $n$-dimensional with components $w_{it|T}^{(k)}$ for $i\in\mathcal I_{m}$ and zero otherwise. 

We also define the $q\times q$ matrices
\begin{align}
&\mbf P_{t|T}^{f(k)}= \E_{\wh{\varphi}_n^{(k)}}\l[(\bm f_{t|T}^{(k)}-\bm f_t)(\bm f_{t|T}^{(k)}-\bm f_{t})^\prime|\bm X_{nT}\r], \nn\\
&\mbf P_{t,t-j|T}^{f(k)}= \E_{\wh{\varphi}_n^{(k)}}\l[(\bm f_{t|T}^{(k)}-\bm f_t)(\bm f_{t-j|T}^{(k)}-\bm f_{t-j})^\prime|\bm X_{nT}\r], \qquad j=1,2,\nn\\
&\mbf P_{t-j,t|T}^{f(k)}= \E_{\wh{\varphi}_n^{(k)}}\l[(\bm f_{t-j|T}^{(k)}-\bm f_{t-j})(\bm f_{t|T}^{(k)}-\bm f_{t})^\prime|\bm X_{nT}\r], \qquad j=1,2,\nn
\end{align}
and the $r\times r$ matrices
\beq
\mbf P_{t|T}^{(k)}=\l(\ba{cc}
\mbf P_{t|T}^{f(k)}&\mbf P_{t,t-1|T}^{f(k)}\\
\mbf P_{t-1,t|T}^{f(k)}&\mbf P_{t|T}^{f(k)}
\ea
\r),\qquad
\mbf P_{t,t-1|T}^{(k)}=\l(\ba{cc}
\mbf P_{t,t-1|T}^{f(k)}&\mbf P_{t,t-2|T}^{f(k)}\\
\mbf P_{t-1|T}^{f(k)}&\mbf P_{t-1,t-2|T}^{f(k)}
\ea
\r)
.\label{eq:PPPPNS}
\eeq
We also define the $n_1\times n_1$ matrices $\mbf P_{t|T}^{1(k)}$ and $\mbf P_{t,t-1|T}^{1(k)}$ with entries
\begin{align}
&[\mbf P_{t|T}^{1(k)}]_{ij} = \E_{\wh{\varphi}_n^{(k)}}\l[( \xi_{it|T}^{(k)}- \xi_{it})( \xi_{jt|T}^{(k)}- \xi_{jt})|\bm X_{nT}\r], \qquad i,j\in\mathcal I_1,\nn\\
&[\mbf P_{t|,t-1T}^{1(k)}]_{ij} = \E_{\wh{\varphi}_n^{(k)}}\l[( \xi_{it|T}^{(k)}- \xi_{it})( \xi_{jt-1|T}^{(k)}- \xi_{jt-1})|\bm X_{nT}\r], \qquad i,j\in\mathcal I_1,\nn
\end{align}
the $n_a\times n_a$ diagonal matrices $\mbf P_{t|T}^{a(k)}$ and $\mbf P_{t,t-1|T}^{a(k)}$ with entries
\begin{align}
&[\mbf P_{t|T}^{a(k)}]_{ii} = \E_{\wh{\varphi}_n^{(k)}}\l[( \alpha_{it|T}^{(k)}- \alpha_{it})^2|\bm X_{nT}\r], \qquad i\in\mathcal I_a,\nn\\
&[\mbf P_{t,t-1|T}^{a(k)}]_{ii} = \E_{\wh{\varphi}_n^{(k)}}\l[( \alpha_{it|T}^{(k)}- \alpha_{it})( \alpha_{it-1|T}^{(k)}- \alpha_{it-1})|\bm X_{nT}\r], \qquad i\in\mathcal I_a,\nn
\end{align}
the $n_b\times n_b$ diagonal matrices $\mbf P_{t|T}^{b(k)}$ and $\mbf P_{t,t-1|T}^{b(k)}$ with entries
\begin{align}
&[\mbf P_{t|T}^{b(k)}]_{ii} = \E_{\wh{\varphi}_n^{(k)}}\l[( \beta_{it|T}^{(k)}- \beta_{it})^2|\bm X_{nT}\r], \qquad i\in\mathcal I_b,\nn\\
&[\mbf P_{t,t-1|T}^{b(k)}]_{ii} = \E_{\wh{\varphi}_n^{(k)}}\l[( \beta_{it|T}^{(k)}- \beta_{it})( \beta_{it-1|T}^{(k)}- \beta_{it-1})|\bm X_{nT}\r], \qquad i\in\mathcal I_b,\nn
\end{align}
and the $\#\mathcal I_m\times \#\mathcal I_m$ diagonal matrix $\mbf P_{t|T}^{w(k)}$  with entries
\begin{align}
&[\mbf P_{t|T}^{w(k)}]_{ii} = \E_{\wh{\varphi}_n^{(k)}}\l[( w_{it|T}^{(k)}- w_{it})^2|\bm X_{nT}\r], \qquad i\in\mathcal I_m.\nn
\end{align}
All those matrices are obtained from the KS. After the first iteration, for any $k\ge 1$ the KF is initialized with $\bm f_{0|0}^{(k)}=\bm f_{0|T}^{(k-1)}$ and
$\mbf P_{0|0}^{(k)}=\mbf P_{0|T}^{(k-1)}$, which is defined as in \eqref{eq:PPPPNS}. \medskip

Denoting as $\underline{\bm\varphi}_n$ the generic values of the parameters, at each iteration $k\ge 0$, the expected log-likelihood is the given by (using the notation of \eqref{practice_obs2})
\beq\label{Estepapp}
\ell(\bm X_{nT};\underline{\bm\varphi}_n) = \E_{\bm\varphi_n^{(k)}}\l[\ell(\bm X_{nT}|\bm S_{nT};\underline{\bm\varphi}_n)\r]+
 \E_{\bm\varphi_n^{(k)}}\l[\ell(\bm S_{nT};\underline{\bm\varphi}_n)\r]- \E_{\bm\varphi_n^{(k)}}\l[\ell(\bm S_{T}|\bm X_{nT};\underline{\bm\varphi}_n)\r]
\eeq
where $\bm X_{nT}$ is the $nT$-dimensional vector containing all data and $\bm S_{nT}$ is the $(r+n_1+n_a+n_b)T$-dimensional vector containing all latent states.
In particular, denoting as $\bm F_T$ the $qT$-dimensional vector containing the $q$ factors, $\bm \Xi_{nT}$ the vector of all $I(1)$ idiosyncratic components, $\bm A_{nT}$ the vector of all time-varying intercepts, and $\bm B_{nT}$ the vector of all time-varying trend slopes, we have
\[
\ell(\bm S_{nT};\underline{\bm\varphi}_n)= \ell(\bm F_{T};\underline{\bm\varphi}_n)+\ell(\bm \Xi_{nT};\underline{\bm\varphi}_n)+\ell(\bm A_{nT};\underline{\bm\varphi}_n)+\ell(\bm B_{nT};\underline{\bm\varphi}_n),
\]
since all groups of states are independent by assumption. Then,
\begin{align}
&\ell(\bm X_{nT}|\bm S_T;\underline{\bm\varphi}_n)\simeq -\frac T2\log\det(\bm\Gamma_n^\nu)- \frac 12\sum_{t=s+1}^T (\mbf x_{nt}-\bm\Upsilon_n\bm s_t)^\prime(\bm\Gamma_n^\nu)^{-1}(\mbf x_{nt}-\bm\Upsilon_n\bm s_t),\nn\\
&\ell(\bm F_{T};\underline{\bm\varphi}_n)\simeq-\frac T2\log\det(\bm\Gamma^u)-\frac 12
\sum_{t=s+1}^T (\bm f_{t}-\bm {\mathcal A_1}\bm f_{t-1}-\bm {\mathcal A_2}\bm f_{t-2})^\prime(\bm\Gamma^u)^{-1}(\bm f_{t}-\bm {\mathcal A_1}\bm f_{t-1}-\bm {\mathcal A_2}\bm f_{t-2}).\nn
\end{align}

\subsection{M-step}
As it is well known, the expected log-likelihood is maximized just by maximizing the first two terms in \eqref{Estepapp}. Therefore, at any iteration $k\ge 0$ of the EM algorithm, we have the following estimators. For the loadings (recall \eqref{eq:PPPPNS}):
\begin{align}
\wh{\bm \lambda}_{i}^{(k+1)}\!=
\l\{\sum_{t=2}^T\l(\ba{ll}
\bm f_{t|T}^{(k)}\bm f_{t|T}^{(k)\prime}&\bm f_{t|T}^{(k)}\bm f_{t-1|T}^{(k)\prime}\\
\bm f_{t-1|T}^{(k)}\bm f_{t|T}^{(k)\prime}&\bm f_{t-1|T}^{(k)}\bm f_{t-1|T}^{(k)\prime}
\ea\r)+\mbf P_{t|T}^{(k)}
\r\}^{\!\!-1}\!\!\!
\l\{\sum_{t=2}^T\l(\ba{l}
\bm f_{t|T}^{(k)}\,x_{it}\\
\bm f_{t-1|T}^{(k)}\,x_{it}
\ea\r)
\r\},\qquad i=1,\ldots, n,\nn
\end{align}
such that, $\wh{\bm b}_{0i}^{(k+1)}$ is given by the first $q$-rows of $\wh{\bm\lambda}^{(k+1)}_i$ and $\wh{\bm b}_{1i}^{(k+1)}$ is given by the other $q$-rows. 

For the VAR parameters:
\begin{align}
\wh{\mbf A}^{(k+1)}\!=\!  \l\{\sum_{t=3}^T
\l(\ba{ll}
\bm f_{t|T}^{(k)}\bm f_{t-1|T}^{(k)\prime}&\bm f_{t|T}^{(k)}\bm f_{t-2|T}^{(k)\prime}\\
\bm f_{t-1|T}^{(k)}\bm f_{t-1|T}^{(k)\prime}&\bm f_{t-1|T}^{(k)}\bm f_{t-2|T}^{(k)\prime}
\ea\r)
+\wh{\mbf P}^{(k)}_{t,t-1|T} \r\}\!
\l\{\sum_{t=3}^T 
\l(\ba{ll}
\bm f_{t-1|T}^{(k)}\bm f_{t-1|T}^{(k)\prime}&\bm f_{t-1|T}^{(k)}\bm f_{t-2|T}^{(k)\prime}\\
\bm f_{t-2|T}^{(k)}\bm f_{t-1|T}^{(k)\prime}&\bm f_{t-2|T}^{(k)}\bm f_{t-2|T}^{(k)\prime}
\ea\r)
+\wh{\mbf P}^{(k)}_{t-1|T}\r\}^{-1}\!\!\!\!,\nn
\end{align}
and, letting, $\wh{\bm {\mathcal A}}_1^{(k+1)}$ be the top-left $q\times q$ block of $\wh{\mbf A}^{(k+1)}$, and $\wh{\bm {\mathcal A}}_2^{(k+1)}$ be the top-right $q\times q$ block of $\wh{\mbf A}^{(k+1)}$, we have
\begin{align}
\wh{\bm\Gamma}^{u(k+1)}=&\;\frac 1 T\sum_{t=3}^T 
\bigg\{
\l({\bm f}^{(k)}_{t|T}{\bm f}^{(k)\prime}_{t|T}+\mbf P_{t|T}^{f(k)}\r)
+\sum_{j=1}^2\bigg[\wh{\bm {\mathcal A}}_j^{(k+1)}
\l(
{\bm f}^{(k)}_{t-j|T}{\bm f}^{(k)\prime}_{t-j|T}+\mbf P_{t-j|T}^{f(k)}
\r)
\wh{\bm {\mathcal A}}_j^{(k+1)\prime}
\nn\\
&-\l(
{\bm f}^{(k)}_{t|T}{\bm f}^{(k)\prime}_{t-j|T}+\mbf P_{t,t-j|T}^{f(k)}
\r)\wh{\bm {\mathcal A}}_j^{(k+1)\prime}
-\wh{\bm {\mathcal A}}_j^{(k+1)}
\l(
{\bm f}^{(k)}_{t-j|T}{\bm f}^{(k)\prime}_{t|T}+\mbf P_{t-j,t|T}^{f(k)}
\r)\bigg]\nn\\
&-\wh{\bm {\mathcal A}}_1^{(k+1)}
\l(
{\bm f}^{(k)}_{t-1|T}{\bm f}^{(k)\prime}_{t-2|T}+\mbf P_{t-1,t-2|T}^{f(k)}
\r)
\wh{\bm {\mathcal A}}_2^{(k+1)\prime}-
\wh{\bm {\mathcal A}}_2^{(k+1)}
\l(
{\bm f}^{(k)}_{t-2|T}{\bm f}^{(k)\prime}_{t-1|T}+\mbf P_{t-2,t-1|T}^{f(k)}
\r)
\wh{\bm {\mathcal A}}_1^{(k+1)\prime}
\bigg\}.\nn
\end{align}
Moreover, the  variances of the state residuals are given by:
\begin{align}
[\wh{\bm\Gamma}_n^e]_{ii} &=\frac 1T\sum_{t=2}^T \bigg\{\xi_{it|T}^{(k)2}+[\mbf P_{t|T}^{1(k)}]_{ii}+\xi_{it-1|T}^{(k)2}+[\mbf P_{t-1|T}^{1(k)}]_{ii}-2\l(\xi_{it|T}^{(k)}\xi_{it-1|T}^{(k)}+[\mbf P_{t,t-1|T}^{1(k)}]_{ii}\r)
\bigg\},\qquad i\in\mathcal I_1,\nn\\
\wh{\sigma}_{i\omega}^{2(k+1)} &=\frac 1T\sum_{t=2}^T \bigg\{\alpha_{it|T}^{(k)2}+[\mbf P_{t|T}^{a(k)}]_{ii}+\alpha_{it-1|T}^{(k)2}+[\mbf P_{t-1|T}^{a(k)}]_{ii}-2\l(\alpha_{it|T}^{(k)}\alpha_{it-1|T}^{(k)}+[\mbf P_{t,t-1|T}^{a(k)}]_{ii}\r)
\bigg\},\qquad i\in\mathcal I_a,\nn\\
\wh{\sigma}_{i\eta}^{2(k+1)} &=\frac 1T\sum_{t=2}^T \bigg\{\beta_{it|T}^{(k)2}+[\mbf P_{t|T}^{b(k)}]_{ii}+\beta_{it-1|T}^{(k)2}+[\mbf P_{t-1|T}^{b(k)}]_{ii}-2\l(\beta_{it|T}^{(k)}\beta_{it-1|T}^{(k)}+[\mbf P_{t,t-1|T}^{b(k)}]_{ii}\r)
\bigg\},\qquad i\in\mathcal I_b,\nn
\end{align}
while the variances of the residuals of the measurement equation are given by
\begin{align}
\wh{\sigma}_{i\nu}^{2(k+1)}=&\;\frac 1 T\sum_{t=2}^T\bigg\{ x_{it}^2 +
\wh{\bm\lambda}_i^{(k+1)\prime}\l(\mbf F_{t|T}^{(k)}\mbf F_{t|T}^{(k)\prime}+\mbf P_{t|T}^{(k)}\r)\wh{\bm\lambda}_i^{(k+1)}+
\l( w_{it|T}^{(k)2}+[\mbf P_{t|T}^{w(k)}]_{ii}\r)\nn\\
&-2x_{it}\l(\wh{\bm\lambda}_i^{(k+1)\prime}\mbf F_{t|T}^{(k)}+w_{it|T}^{(k)}\r)
-2 \wh{\bm\lambda}_i^{(k+1)\prime}\mbf F_{t|T}^{(k)} w_{it|T}^{(k)}
\bigg\},\qquad i\in\mathcal I_m,\nn\\
\wh{\sigma}_{i\nu}^{2(k+1)}\equiv&\;
[\wh{\bm\Gamma}_n^e]_{ii} =\frac 1 T\sum_{t=2}^T\bigg\{ x_{it}^2 +
\wh{\bm\lambda}_i^{(k+1)\prime}\l(\mbf F_{t|T}^{(k)}\mbf F_{t|T}^{(k)\prime}+\mbf P_{t|T}^{(k)}\r)\wh{\bm\lambda}_i^{(k+1)}-2x_{it}\wh{\bm\lambda}_i^{(k+1)\prime}\mbf F_{t|T}^{(k)}
\bigg\},\qquad i\notin\mathcal I_m.\nn
\end{align}
Finally, we set $[\wh{\bm\Gamma}_n^e]_{ij} =0$ for all $i,j=1,\ldots,n$ such that $i\ne j$.
\clearpage

%
%
\section{Proof of Proposition \ref{th:chiNS}}\label{app:consN}
The proof follows the same steps as the proof of consistency in Theorem 1
in \citet{BLqml}, and unless substantial differences emerge, we refer to results therein for detailed proofs of all the intermediate steps.

Throughout, for simplicity, and without loss of generality, we let $s=1$, so that $r=q(s+1)=2q$, and we let also $p=2$. Recall also that we are considering the case in which $n_1=0$, $n_a=0$, and $n_b=0$.\\

\noindent
\underline{\it Stabilizability and detectability.}  Recall the state space form \eqref{eq:SSNS001}-\eqref{eq:SSNS002} of the NS-DFM
\begin{align}
\mbf x_{nt}&=\l(\bm{\mathcal B}_{0n}\ \bm{\mathcal B}_{1n}\r)
\l(\ba{c}
\bm f_t\\
\bm f_{t-1}
\ea
\r)+\mbf e_{nt},\label{eq:SSNS1}\\
\l(\ba{c}
\bm f_t\\
\bm f_{t-1}
\ea
\r)&=\l(\ba{cc}
\bm{\mathcal A}_1&\bm{\mathcal A}_2\\
\mbf I_q&\mbf 0_{q\times q}
\ea
\r)
\l(\ba{c}
\bm f_{t-1}\\
\bm f_{t-2}
\ea
\r)
+\l(\ba{c}
\bm u_t\\
\mbf 0_q
\ea
\r),\label{eq:SSNS2}
\end{align} 
%
%
Then, \eqref{eq:SSNS1}-\eqref{eq:SSNS2} define a linear system with $r=2q$ latent states $(\bm f_t'\;\bm f_{t-1}')'$. 

A linear system is stabilizable if its unstable states are controllable and all uncontrollable states are stable, and it is detectable if its unstable states are observable and all unobservable states are stable (see \citealp[Appendix C, page 342]{AM79}). 

Let us first show that \eqref{eq:SSNS1}-\eqref{eq:SSNS2} is stabilizable. Stability is dictated by the eigenvalues of the matrix of VAR coefficients, 
\begin{align}
{\mbf A}=\l(\ba{cc}
\bm{\mathcal A}_1&\bm{\mathcal A}_2\\
\mbf I_q&\mbf 0_{q\times q}
\ea
\r).\label{eq:Asingolare}
\end{align}
Because of cointegration, ${\mbf A}$ has $(q-d)$ unit eigenvalues corresponding to $(q-d)$ unstable states. Moreover, $(\mbf I_q-\bm{\mathcal A}_1-\bm{\mathcal A}_2)=\mbf a\mbf b'$, where $\mbf a$ and $\mbf b$ have full column-rank $q\times d$ matrices, so that $\text{rk}(\mbf a\mbf b')=d$. Define the $q\times (q-d)$ matrices $\mbf a_{\perp}$ and $\mbf b_{\perp}$ such that
$\mbf a_{\perp}'\mbf a=\mbf b_{\perp}'\mbf b=\mbf 0_{(q-d)\times d}$. Then, since $\text{rk}(\mbf a_{\perp}'\mbf I_q)=(q-d)$, the unstable states are controllable because they satisfy the Popov-Belevitch-Hautus rank test (see \citealp{franchi}, Theorem 2.1, and \citealp{AM07}, Corollary 6.11, page 249). Clearly, ${\mbf A}$ has also $(r-q+d)=(q+d)$ eigenvalues which are smaller than one in absolute value. Of these $q$ correspond to states which are uncontrollable because they are not driven by any shock, but are also stable since have no dynamics (see \eqref{eq:SSNS2}). The remaining $d$ states follow a stable VAR, hence are controllable.

Let us now show that \eqref{eq:SSNS1}-\eqref{eq:SSNS2} is detectable. First, notice that $\text{rk}(\bm{\mathcal B}_{0n})=q$ and $\text{rk}(\bm{\mathcal B}_{1n})=q$, because of Assumption \ref{ass:dynamic}(a) and we are assuming pervasive factors at all lags. Therefore, $\text{rk}(\bm{\mathcal B}_{0n} \mbf b_{\perp})=(q-d)$ and $\text{rk}(\bm{\mathcal B}_{1n} \mbf b_{\perp})=(q-d)$, which implies that the unstable states are observable because they satisfy the Popov-Belevitch-Hautus rank test (see \citealp{franchi}, Theorem 2.1, and \citealp{AM07}, Corollary 6.11, page 249). Since $\bm{\mathcal B}_{0n}$ and $\bm{\mathcal B}_{1n}$ have full column-rank there are no unstable unobservable states. \\

\noindent
\underline{\it Estimation of factors given parameters.}  For the linear system in \eqref{eq:SSNS1}-\eqref{eq:SSNS2}, define 
\beq
\bm\Lambda_n=(\bm{\mathcal B}_{0n}\, \bm{\mathcal B}_{1n}),\qquad \mbf F_t=(\bm f_t^\prime\ \bm f_{t-1}^\prime)^\prime.\label{eq:Bsingolare}
\eeq 
Then, using the definitions in \eqref{eq:Asingolare} and \eqref{eq:Bsingolare} and by setting $\mbf K=\mbf I_r$, the results in Lemmas 
4, 5, and 6 of \citet{BLqml}, still hold.
In particular, since the system is stabilizable and detectable, the matrix $\mbf P_{t|t-1}$ has a steady state denoted as ${\mbf P}$, and there exists a positive integer $\bar n$, such that, for any $n\ge \bar n$,
$$
\l\Vert{\mbf P} - \l(\ba{cc}
\mbf I_q & \mbf 0_{q\times q}\\
\mbf 0_{q\times q}&\mbf 0_{q\times q}
\ea
\r)
\r\Vert \le Mn^{-1},
$$
for some positive real $M$. Moreover, notice that in the proof of Lemma 6 of \citet{BLqml} it is enough that $\Vert \mbf A \Vert\le 1$, which is always satisfied because of Assumption \ref{ass:dynamic}(e). The definition of $\bar t$ is also unchanged.

Consistency can then be proved as in Proposition 1
of \citet{BLqml}. By letting $\bm f_{t|T}$ be the KS estimate of $\bm f_t$ (given by the first $q$ components of $\mbf F_{t|T}$), as $n\to \infty$, for any given $t\ge \bar t$, we have
\beq
\sqrt n\Vert \bm f_{t|T}-\bm f_t\Vert = O_p(1).\label{PRIMO}
\eeq
This proves the analogous of Proposition 1 of \citet{BLqml} for the NS-DFM.\\

\noindent
\underline{\it QML estimation of parameters given factors.} Recalling the definitions \eqref{eq:Asingolare} and \eqref{eq:Bsingolare}, the QML estimator of the loadings, for any $i=1,\ldots,n$, is given by
\begin{align}
\wh{\bm\lambda}_i^{*}=
\l(\sum_{t=1}^T {\mbf F}_{t}{\mbf F}_{t}^\prime\r)^{-1}\l(\sum_{t=1}^T   {\mbf F}_{t}x_{it}\r).\label{DAXX1}
\end{align}
Because of Assumption \ref{ass:dynamic}(f), $\mbf F_t$ is cointegrated and admits a common trends representation with $(q-d)$ common trends \citep{stockwatson88JASA}. Therefore, we can find an orthonormal linear basis of dimension $(q-d)$ such that the projection of $\mbf F_t$ onto this basis span the same space as the common trends. Collect the elements of this basis in the $r\times (q-d)$ matrix $\bm\gamma$, and denote as $\bm\gamma_\perp$ the $r\times(r-q+ d)$ matrix such that $\bm\gamma_\perp^\prime\bm\gamma=\mbf 0_{(r-q+d)\times (q-d)}$. Then, consider the $r\times r$ linear  transformation 
\beq\label{DFZ}
\bm{\mathcal D}\mbf F_t = \l(\ba{c}
\bm\gamma^\prime\\
\bm\gamma_\perp^\prime\\
\ea
\r)\mbf F_t=\l(\ba{c}
\mbf Z_{1t}\\
\mbf Z_{0t}
\ea\r),\; \text{say},
\eeq
where $\mbf Z_{1t}$ has all $(q-d)$ components which are $I(1)$ while $\mbf Z_{0t}\sim I(0)$ and is of dimension $(r-q+d)$. Moreover, for $\mbf Z_{1t}$ we have the MA representation  
\beq\label{z1Q}
\Delta \mbf Z_{1t} = \sum_{k=0}^\infty\mbf Q_k \bm z_{t-k},
\eeq
with $\bm z_t$ being a $(r-q+d)$-dimensional vector with $\E_{\varphi_n}[\bm z_t]=\mbf 0_{(q-d)}$, $\E_{\varphi_n}[\bm z_t\bm z_t^\prime]=\bm\Sigma_{z}$ positive definite and with finite norm, and $\E_{\varphi_n}[\bm z_s\bm z_t^\prime]=\mbf 0_{(q-d)\times (q-d)}$, for any $s\ne t$. Moreover, $\text{rk}({\mbf Q}(1))=(q-d)$, and $\sum_{k=0}^\infty\Vert \mbf Q_k\Vert^2<\infty$. 
 
Because of orthonormality $\bm{\mathcal D}^\prime\bm{\mathcal D}=\mbf I_r$. Then, let $\bm\lambda_i^\prime\bm{\mathcal D}^\prime=(\bm\lambda_{i1}^\prime\; \bm\lambda_{i0}^\prime)$ such that \eqref{eq:NSDFM1} reads (recall we are considering the case $\xi_{it}=e_{it}$),
\beq\label{DAXX}
x_{it}=\bm\lambda_{i1}^\prime\mbf Z_{1t}+\bm\lambda_{i0}^\prime\mbf Z_{0t}+e_{it},
\eeq 
and define also $\wh{\bm\lambda}_i^{*\prime}\bm{\mathcal D}^\prime=(\wh{\bm\lambda}_{i1}^{*\prime}\;\wh{\bm\lambda}_{i0}^{*\prime})$. Since by construction $\mbf Z_{1t}\mbf Z_{0t}^\prime=\mbf 0_{(q-d)\times(r-q+d)}$ and $\mbf Z_{0t}\mbf Z_{1t}^\prime=\mbf 0_{(r-q+d)\times(q-d)}$, from \eqref{DAXX1} and \eqref{DAXX}, we have
\begin{align}
&\l(\ba{c}
\wh{\bm\lambda}_{i1}^{*}-\bm\lambda_{i1}\\
\wh{\bm\lambda}_{i0}^{*}-\bm\lambda_{i0}
\ea\r)\label{lambdaDD}\\
&=\l(\ba{cc}
\l(T^{-2}\sum_{t=1}^T {\mbf Z}_{1t}{\mbf Z}_{1t}^\prime\r)^{-1}\l(T^{-2} \sum_{t=1}^T   {\mbf Z}_{1t}e_{it}\r)&\mbf 0_{(q-d)\times (r-q+d)}\\
\mbf 0_{(r-q+d)\times (q-d)}&
\l(T^{-1}\sum_{t=1}^T {\mbf Z}_{0t}{\mbf Z}_{0t}^\prime \r)^{-1}\l(T^{-1} \sum_{t=1}^T  {\mbf Z}_{0t}^\prime e_{it}\r)
\ea
\r).\nn
\end{align}
First, consider the top left term on the rhs of \eqref{lambdaDD}. From \citet[Proposition 18.1(i) pages 547-548]{Hamilton} and \eqref{z1Q}, as $T\to\infty$,
\beq
T^{-2}\sum_{t=1}^T {\mbf Z}_{1t}{\mbf Z}_{1t}^\prime\stackrel{d}{\to} \mbf Q(1)\bm\Sigma_{z}^{1/2}\l(\int_0^1\bm{\mathcal W}(u)\bm{\mathcal W}(u)^\prime\mathrm d u\right)\bm\Sigma_{z}^{1/2}\mbf Q(1)^\prime,\label{PPAPP}
\eeq
where $\bm{\mathcal W}(\cdot)$ is a $(q-d)$-dimensional standard Wiener process. Thus this term is $O_p(1)$ and positive definite therefore invertible. Furthermore, for all $i=1,\ldots, (q-d)$ and all $t=1,\ldots, T$,
\beq\label{eq:varrw}
\E_{\varphi_n}[Z_{1it}^2] =\sum_{s=1}^t\sum_{k=0}^\infty \sum_{j,\ell=1}^{(q-d)} [\mbf Q_k]_{ij}[\bm\Sigma_z]_{j\ell} [\mbf Q_k]_{\ell i}\le C_i t,
\eeq
for some positive real $C_i$ and because of square summability of the MA coefficients in \eqref{z1Q} and since $\bm\Sigma_z$ has finite norm. 
Thus, since $\mbf F_t$ and $e_{it}$ are gaussian and uncorrelated by Assumptions \ref{ass:modelNS}(a), \ref{ass:modelNS}(b), and \ref{ass:modelNS}(d), then they are also mutually independent, and we have 
\begin{align}\label{casino}
\E_{\varphi_n}\l[\l\Vert T^{-2}\sum_{t=1}^T {\mbf Z}_{1t} e_{it}\r\Vert^2\r] &= T^{-4}\sum_{j=1}^{(q-d)}\sum_{t,s=1}^T\E_{\varphi_n}[Z_{1jt}Z_{1js} z_{it} z_{is}]=T^{-4}\sum_{j=1}^{(q-d)}\sum_{t,s=1}^T\E_{\varphi_n}[Z_{1jt}Z_{1js}] \E_{\varphi_n}[z_{it} z_{is}]\nn\\
&=T^{-4}\sum_{j=1}^{(q-d)}\sum_{t=1}^T\E_{\varphi_n}[Z_{1jt}^2] \E_{\varphi_n}[z_{it}^2]\le T^{-4} \sum_{j=1}^{(q-d)} C_j \sum_{t=1}^T t \E_{\varphi_n}[z_{it}^2]= O\l({T}^{-2}\r),
\end{align}
where we used \eqref{eq:varrw} and the fact that $\bm z_t$ is a white noise process with finite variance. From, \eqref{PPAPP} and \eqref{casino}, we have
\beq
\Vert \wh{\bm\lambda}_{i1}^{*}-\bm\lambda_{i1}\Vert = O_p(T^{-1}).\label{lambda1cons}
\eeq
Second, consider the bottom right term on the rhs of \eqref{lambdaDD}. By the same arguments used to prove Lemma 
8(i) in \citet{BLqml}, we have
\begin{align}
T^{-1} \sum_{t=1}^T  {\mbf Z}_{0t} e_{it} = O_p(T^{-1/2}),\qquad T^{-1} \sum_{t=1}^T  \mbf Z_{0t} {\mbf Z}_{0t}' = O_p(1),\label{Z0Z0}
\end{align}
which imply
\beq
\Vert \wh{\bm\lambda}_{i0}^{*}-\bm\lambda_{i0}\Vert = O_p(T^{-1/2}).\label{lambda0cons}
\eeq
By substituting \eqref{lambda1cons} and \eqref{lambda0cons} into \eqref{lambdaDD} and since  since $\bm{\mathcal D}$ does not depend on $T$, as $T\to\infty$, for any given $i=1,\ldots,n$, we  have
\beq
\sqrt T\Vert \wh{\bm\lambda}_{i}^{*}-\bm\lambda_{i}\Vert = O_p(1).\label{SECONDO}
\eeq

Turning to estimation of the VAR coefficients, the QML estimator is given by
\begin{align}
\wh{\mbf A}^{*}=
\l(\sum_{t=2}^T   {\mbf F}_{t}\mbf F_{t-1}^\prime\r)\l(\sum_{t=2}^T {\mbf F}_{t}{\mbf F}_{t}^\prime\r)^{-1}.\label{DADFF1}
\end{align}
From \eqref{eq:NSDFM2}, we can also write
\beq
\bm{\mathcal D}\mbf F_t = (\bm{\mathcal D}\mbf A\bm{\mathcal D}^\prime)\bm{\mathcal D}\mbf F_{t-1}+\bm{\mathcal D}\bm u_t.\label{DADFF}
\eeq
such that $\bm{\mathcal D}\bm u_t=(\mbf v_{1t}^\prime\;\mbf {v}_{0t}^\prime)^\prime$ where $\mbf v_{1t}$ and $\mbf v_{0t}$ are zero mean white noise processes of dimensions $(q-d)$ and $(r-q+d)$, respectively. Then, similarly to \eqref{lambdaDD}, from \eqref{DADFF1} and \eqref{DADFF}, we have
\begin{align}
&\bm{\mathcal D}(\wh{\mbf A}^*-\mbf A)\bm{\mathcal D}^\prime=\label{DAD}\\
&=\l(\ba{cc}
\l(T^{-2} \sum_{t=2}^T  \mbf v_{1t} {\mbf Z}_{1t-1}^\prime\r)
\l(T^{-2}\sum_{t=2}^T {\mbf Z}_{1t-1}{\mbf Z}_{1t-1}^\prime\r)^{-1}&
\l(T^{-2} \sum_{t=2}^T  \mbf v_{0t} {\mbf Z}_{1t-1}^\prime\r)
\l(T^{-2}\sum_{t=2}^T {\mbf Z}_{1t-1}{\mbf Z}_{1t-1}^\prime\r)^{-1}\\
\l(T^{-1} \sum_{t=2}^T  \mbf v_{1t} {\mbf Z}_{0t-1}^\prime\r)
\l(T^{-1}\sum_{t=2}^T {\mbf Z}_{0t-1}{\mbf Z}_{0t-1}^\prime\r)^{-1}&
\l(T^{-1} \sum_{t=2}^T  \mbf v_{0t} {\mbf Z}_{0t-1}^\prime\r)
\l(T^{-1}\sum_{t=2}^T {\mbf Z}_{0t-1}{\mbf Z}_{0t-1}^\prime\r)^{-1}
\ea
\r).\nn
\end{align}
Then, using the fact that $\mbf v_{1t}$ and $\mbf v_{0t}$ are gaussian white noise and therefore are martingale difference sequences, from \citet[Proposition 18.1(f) pages 547-548]{Hamilton}, it follows that
\begin{align}
&T^{-2} \sum_{t=2}^T  \mbf v_{1t} {\mbf Z}_{1t-1}^\prime = O_p(T^{-1}), && T^{-2}\sum_{t=2}^T  \mbf v_{0t} {\mbf Z}_{1t-1}^\prime = O_p(T^{-1}),\label{altriOP}
\end{align}
and, from \citet[Proposition 11.1, pages 298-299]{Hamilton}, it follows that
\begin{align}
&T^{-1} \sum_{t=2}^T  \mbf v_{1t} {\mbf Z}_{0t-1}^\prime = O_p(T^{-1/2}),&&T^{-1} \sum_{t=2}^T  \mbf v_{0t} {\mbf Z}_{0t-1}^\prime = O_p(T^{-1/2}).\label{altriOPbis}
\end{align}
Substituting \eqref{PPAPP}, \eqref{Z0Z0}, \eqref{altriOP} and \eqref{altriOPbis} into \eqref{DAD}, and since $\bm{\mathcal D}$ does not depend on $T$, we have 
\beq
\Vert \wh{\mbf A}^{*}-\mbf A\Vert = O_p(T^{-1/2}).
\eeq
Finally,  the process $T^{-2}\Vert\sum_{t=1}^T \mbf Z_{1t} e_{it}\Vert$ is gaussian, with zero-mean and variance $O(T^{-2})$. Then, using Bonferroni inequality, and noticing that the rhs of \eqref{casino} does not depend on $i$, there exists a finite positive real $K_1$, independent of $i$, such that  for all $\epsilon >0$
\beq
\text{P}_{\varphi_n}\l( \max_{1\le i\le n} T^{-2} \l\Vert \sum_{t=1}^T \mbf Z_{1t} e_{it}\r\Vert >\epsilon\r)\le n \max_{1\le i\le n} \text{P}_{\varphi_n}\l( T^{-2} \l\Vert \sum_{t=1}^T \mbf Z_{1t} e_{it}\r\Vert >\epsilon\r)\le n \exp(-K_1T^2\epsilon^2).\label{eq:unifload1}
\eeq 
and, similarly, there exists a finite positive real $K_0$, independent of $i$, such that  for all $\epsilon >0$
\beq
\text{P}_{\varphi_n}\l( \max_{1\le i\le n} T^{-1} \l\Vert \sum_{t=1}^T \mbf Z_{0t} e_{it}\r\Vert >\epsilon\r)\le n \max_{1\le i\le n} \text{P}_{\varphi_n}\l( T^{-1} \l\Vert \sum_{t=1}^T \mbf Z_{0t} e_{it}\r\Vert >\epsilon\r)\le n \exp(-K_0T\epsilon^2).\label{eq:unifload0}
\eeq 
From \eqref{eq:unifload1}, \eqref{eq:unifload0} and \eqref{SECONDO}, $\max_{i=1,\ldots,n}\sqrt T\Vert \wh{\bm\lambda}_{i}^{*}-\bm\lambda_{i}\Vert = O_p(\sqrt{\log n})$. Then, for the estimator $\wh{\bm\Gamma}_n^{e*}$  of $\bm\Gamma_n^e$ the same consistency proof given in Lemma 8(ii)
in \citet{BLqml} still holds.  This proves the analogous of Lemma 8 in \citet{BLqml} for the NS-DFM.\\

\noindent
\underline{\it Estimation of factors given QML estimates of parameters.} 
First, notice that using the notation of \eqref{DFZ} 
we have $\Vert\mbf Z_{1t}\Vert=O_p(\sqrt T)$ and $\Vert\mbf Z_{0t}\Vert=O_p(1)$. 
Then, the same steps leading to the proof of Lemma 9
in \citet{BLqml} still hold, where, in particular, 
we can make use of the following relations (recall that $\bm{\mathcal D}^\prime\bm{\mathcal D}=\mbf I_r$):
\begin{align}
&(\wh{\bm{\lambda}}_i^{*\prime}-{\bm{\lambda}}_i^\prime)\mbf A\mbf F_t=(\wh{\bm{\lambda}}_i^{*\prime}-{\bm{\lambda}}_i^\prime)\bm{\mathcal D}^\prime\bm{\mathcal D}\mbf A\bm{\mathcal D}^\prime\bm{\mathcal D}\mbf F_t=O_p(T^{-1/2}), \label{DF2}\\
&(\wh{\mbf A}^*-\mbf A)\mbf F_t=\bm{\mathcal D}'\bm{\mathcal D}(\wh{\mbf A}^*-\mbf A)\bm{\mathcal D}'\bm{\mathcal D}\mbf F_t=O_p(T^{-1/2}),\label{DF1}
\end{align} 
where \eqref{DF2} holds because of \eqref{lambdaDD}, \eqref{lambda1cons} and \eqref{lambda0cons}, and, similarly, \eqref{DF1} holds because of \eqref{DAD}, \eqref{altriOP} and \eqref{altriOPbis}. Therefore, as $n,T\to\infty$, for any given $t\ge \bar t$, we have
\beq\label{TERZO}
\min(\sqrt n,\sqrt T)\Vert \bm f_{t|T}^*-\bm f_{t}\Vert= O_p(1).
\eeq
This proves  the analogous of Lemma 9
in \citet{BLqml} for the NS-DFM.\\

\noindent
\underline{\it Consistency of pre-estimator of parameters.} Under Assumption \ref{ass:identNS}, the pre-estimators defined in Section \ref{app:prestNS} are such that, for any given $i=1,\ldots, n$,
\begin{align}
&\Vert\wh{\bm{b}}_{0i}^{(0)}-\bm{ b}_{0i}\Vert= O_p(\max(n^{-1/2},T^{-1/2})),\qquad  \Vert \Delta\wt{\bm f}_t-\Delta\bm f_t\Vert = O_p(\max(n^{-1/2},T^{-1/2})),\label{eq:consftildeNS},
\end{align}
see \citet[Lemma 3]{BLL2} and also \citet[Lemma 1]{baing04}. Moreover, it is easy to show that
\beq
\l \Vert T^{-1}\sum_{t=2}^T\Delta\wt{\bm f}_{t-1}\Delta\wt{\bm f}_{t-1}^\prime - T^{-1}\sum_{t=2}^T\Delta{\bm f}_{t-1}\Delta{\bm f}_{t-1}^\prime\r\Vert= O_p(\max(n^{-1/2},T^{-1/2})).\label{eq:consftildeNS2}
\eeq
Then,
\begin{align}
\Vert\wh{\bm{b}}_{1i}^{(0)}-\bm{ b}_{1i}\Vert\le&\,\l\Vert \l(T^{-1} \sum_{t=2}^T\Delta\wt{\bm f}_{t-1}\Delta\wt{\bm f}_{t-1}^\prime\r)^{-1}
\l(T^{-1}\sum_{t=2}^T \Delta\wt{\bm f}_{t-1}(\bm{ b}_{0i}^\prime\Delta\bm f_t-\wh{\bm{b}}_{0i}^{(0)\prime}\Delta\wt{\bm f}_t)\r)\r\Vert\nn\\
&+\l\Vert
\l(T^{-1} \sum_{t=2}^T\Delta\wt{\bm f}_{t-1}\Delta\wt{\bm f}_{t-1}^\prime\r)^{-1}
\l(T^{-1}\sum_{t=2}^T \Delta\wt{\bm f}_{t-1}\Delta e_{it}\r)\r\Vert\nn\\
&+\l\Vert\l(T^{-1} \sum_{t=2}^T\Delta\wt{\bm f}_{t-1}\Delta\wt{\bm f}_{t-1}^\prime\r)^{-1}\l(T^{-1}\sum_{t=2}^T \bm b_{1i}^\prime \Delta\bm f_{t-1}\Delta\wt{\bm f}_{t-1}\Delta e_{it}\r)-\bm b_{1i}^\prime\r\Vert\nn\\
=&\,O_p(\max(n^{-1/2},T^{-1/2})),
\end{align}
which follows from \eqref{eq:consftildeNS} and  \eqref{eq:consftildeNS2}, and noticing that, since $\Delta\bm f_{t-1}$ and $\Delta e_{it}$ are gaussian and uncorrelated by Assumptions \ref{ass:modelNS}(a), \ref{ass:modelNS}(b), and \ref{ass:modelNS}(d), then they are also mutually independent, and therefore
\begin{align}
\E_{\varphi_n}\l[\l \Vert T^{-1}\sum_{t=2}^T\Delta{\bm f}_{t-1}\Delta e_{it}\r\Vert^2\r]= T^{-2} \sum_{j=1}^r \sum_{s,t=2}^T \E_{\varphi_n}[\Delta{f}_{js-1}\Delta{f}_{jt-1}]\E_{\varphi_n}[\Delta e_{is}\Delta e_{it}]=O(T^{-1}),\nn
\end{align}
since $\E_{\varphi_n}[\Delta e_{is}\Delta e_{it}]>0$ only if $|t-s|\le 1$ because $\Delta e_{it}$ is an MA(1). For the same reason $\Var_{\varphi_n}[\Delta e_{it}]=2[\bm\Gamma_n^e]_{ii}$, thus from \eqref{eq:consftildeNS} we have consistency of the diagonal elements of $\wh{\bm\Gamma}_n^{e(0)}$, while for the off-diagonal terms 
\[
n^{-2}\sum_{\substack {i,j=1\\i\ne j} }^n\gamma_{ij}^2 \le n^{-2} \Vert \bm\Gamma_n^e\Vert_F^2 = n^{-2}\mbox{tr}(\bm\Gamma_n^e\bm\Gamma_n^e)
\le n^{-1} \nu^{(1)}(\bm\Gamma_n^e) = n^{-1}\Vert \bm\Gamma_n^e\Vert^2\le n^{-1} \Vert \bm\Gamma_n^e\Vert_1^2\le n^{-1}M_e^2,
\]
because of Assumption \ref{ass:modelNS}(c). 

Last, $\Vert\wh{\mbf A}^{(0)}-\mbf A\Vert= O_p(\max(n^{-1/2},T^{-1/2}))$, because of \citet[Proposition 2]{BLL2}. This proves the analogous of Lemma 10
in \citet{BLqml} for the NS-DFM.\\

\noindent
\underline{\it Convergence of EM estimator.} The proof of Lemma 11
parts (i), (iii), and (iv) in \citet{BLqml} holds also in the NS-DFM with no modifications. Thus, as $n,T\to\infty$, for any given $i=1,\ldots, n$,
\beq\label{QUARTO}
\min(\sqrt n,\sqrt T)\Vert \wh{\bm b}_{0i}-\bm b_{0i}\Vert = O_p(1), \qquad \min(\sqrt n,\sqrt T)\Vert \wh{\bm b}_{1i}-\bm b_{1i}\Vert = O_p(1).
\eeq
From \eqref{QUARTO}, using the same reasoning leading to \eqref{TERZO}, we also have, as $n,T\to\infty$, for any given $t=\bar t,\ldots, T$,
\beq\label{QUINTO}
\min(\sqrt n,\sqrt T)\Vert \wh{\bm f}_{t|T}-\bm f_{t}\Vert= O_p(1).
\eeq
Therefore, from \eqref{QUARTO} and \eqref{QUINTO}, as $n,T\to\infty$, for any given $i=1,\ldots,n$ and $t=(\bar t+1),\ldots, T$,
\[
\min(\sqrt n,\sqrt T)\vert \wh{\chi}_{it}-\chi_{it}\vert=O_p(1).
\]
This proves Proposition \ref{th:chiNS}. $\Box$

\section{The case of additional latent states}\label{app:misidioNS}
For simplicity assume a static one factor model, thus $q=1$, $d=0$ and $s=0$, and also assume that $n_1=m$ while $n_a=0$ and $n_b=0$.
Moreover, for any given $n\in\mathbb N$, assume that we always order the variables in such a way that $\mathcal I_m=\{1,\ldots,m\}$. Furthermore, define the $m\times 1$ vector $\mbf e_{1t}=(e_{1t}\cdots e_{mt})^\prime$ with covariance matrix $\bm\Gamma_1^e$, and the $(n-m)\times 1$ vector $\mbf e_{0t}=(e_{m+1t}\cdots e_{nt})^\prime$ with covariance matrix $\bm\Gamma_0^e$. Notice that $\bm\Gamma_1^e$ and $\bm\Gamma_0^e$ still satisfy Assumptions \ref{ass:modelNS}(b) and  \ref{ass:modelNS}(c).

Recall also that $\E_{\varphi_n}[e_{it}u_{t}]=0$ for all $i=1,\ldots, n$, because of Assumption \ref{ass:modelNS}(d). Then, partition the $n\times 1$ loadings vector as $\bm{\mathcal B}_n=(\bm{\mathcal B}_{1}^\prime\ \bm{\mathcal B}_{0}^\prime)$, where $\bm{\mathcal B}_{1}=(\lambda_1\cdots \lambda_m)^\prime$ is $m\times 1$, and $\bm{\mathcal B}_{0}=(\lambda_{m+1}\cdots \lambda_n)^\prime$ is $(n-m)\times 1$. Consistently with Assumption \ref{ass:static}(a) let us assume that $\lim_{m\to\infty}m^{-1}\bm{\mathcal B}_{1}^\prime\bm{\mathcal B}_{1}=1$ and $\lim_{n,m\to\infty}(n-m)^{-1}\bm{\mathcal B}_{0}^\prime\bm{\mathcal B}_{0}=1$. Let $\bm \xi_{t}=(\xi_{1t}\cdots \xi_{mt})^\prime$, while $\xi_{it}=e_{it}$ for $i=m+1,\ldots, n$, and $\bm\nu_t=(\nu_{1t}\cdots\nu_{mt})^\prime$. 
Last, let $\bm P=\mbox{diag}(\rho_1\cdots \rho_n)$. To avoid heavy notation we omit the dependence on $n$ and $m$ of the matrices and vectors considered.

We have the state space form
\begin{align}
\mbf x_{t}\equiv\l(\ba{c}
\mbf x_{1t}\\
\mbf x_{0t}
\ea
\r)&=
\l(
\ba{cc}
\mbf I_m &\bm{\mathcal B}_1\\
\mbf 0_{(n-m)\times m}&\bm{\mathcal B}_0
\ea
\r)\l(\ba{c}
\bm \xi_{t}\\
f_t
\ea
\r)+
\l(\ba{c}
\bm \nu_{t}\\
\mbf e_{0t}
\ea
\r), \label{eq:ssidiocorr}\\
\l(\ba{c}
\bm \xi_{t}\\
f_t
\ea
\r)&= \l(\ba{cc}
\bm P&\mbf 0_{m\times 1}\\
\mbf 0_{1\times m}& 1
\ea
\r)\l(\ba{c}
\bm \xi_{t-1}\\
f_{t-1}
\ea
\r)+\l(\ba{c}
\mbf e_{1t}\\
u_t
\ea
\r).\nn
\end{align}
Moreover, the error terms in \eqref{eq:ssidiocorr} are such that
\begin{align}
& \l(\ba{c}
\bm \nu_{t}\\
\mbf e_{0t}
\ea
\r)\sim \mathcal N\l(
\l(\ba{c}
\mbf 0_m\\
\mbf 0_{(n-m)}
\ea
\r),
\l(\ba{cc}
\phi \mbf I_m&\mbf 0_{m\times (n-m)}\\
\mbf 0_{(n-m)\times m}& \bm\Gamma_{0}^e
\ea
\r)
\r)
\nn\\
& \l(\ba{c}
\mbf e_{1t}\\
u_{t}
\ea
\r)\sim \mathcal N\l(
\l(\ba{c}
\mbf 0_m\\
0
\ea
\r),
\l(\ba{cc}
\bm\Gamma_1^e&\mbf 0_{m\times 1}\\
\mbf 0_{1\times m}& \sigma_u^2
\ea
\r)
\r),\nn
\end{align}
where $\sigma_u^2=\E_{\varphi_n}[u_{t}^2]$. Notice that $\bm\nu_t$, $\mbf e_{0t}$, $\mbf e_{1t}$, and $u_t$ are all white noise processes. Denote the $(m+1)$-dimensional state vector as $\bm s_t = (\bm\xi_t^\prime \ f_t)^\prime$, such that $\bm S_T=(\bm s_1^\prime\cdots \bm s_T^\prime)^\prime$.  The parameter vector becomes  $\bm\varphi_n=(\text{vec}(\bm{\mathcal B}_n)^\prime, \text{vech}(\bm\Gamma_n^e)^\prime, \rho_1,\ldots, \rho_n ,\text{vec}(\mbf A)^\prime,\text{vec}(\mbf H)^\prime)^\prime$. Notice that as shown later we need to control $\phi$ exogenously in order to achieve consistency, hence here we consider  $\phi$ as given.

The log-likelihood of the data given the factors and the idiosyncratic states and for generic values of the parameters, $\underline{\bm\varphi}_n$, is
\begin{align}
\ell(\bm X_{nT}|\bm S_T;\underline{\bm\varphi}_n)\simeq&\, -\frac T2\log\det(\underline{\bm\Gamma}_0^e)-\frac 12\sum_{t=1}^T(\mbf x_{1t}-\bm \xi_t-\underline{\bm{\mathcal B}}_1 f_t)^\prime \phi^{-1}(\mbf x_{1t}-\bm \xi_t-\underline{\bm{\mathcal B}}_1 f_t)\nn\\
&-\frac 12\sum_{t=1}^T(\mbf x_{0t}-\underline{\bm{\mathcal B}}_0 f_t)^\prime (\bm\Gamma_0^e)^{-1}(\mbf x_{0t}-\underline{\bm{\mathcal B}}_0 f_t).\label{eq:nuovaLLL}
\end{align}
By maximizing \eqref{eq:nuovaLLL} we have the QML estimator of the loadings:
\begin{align}
&\wh{\bm{\mathcal B}}_{1}^{*} = \l(\sum_{t=1}^T (\mbf x_{1t}-\bm\xi_{1t}){f}_{t} \r)
\l(\sum_{t=1}^T f_t^2 \r)^{-1},\qquad \wh{\bm{\mathcal B}}_{0}^{*} = \l(\sum_{t=1}^T \mbf x_{0t}{f}_{t} \r)
\l(\sum_{t=1}^T f_t^2 \r)^{-1}.\nn
\end{align}
The formulas for the estimators obtained  in the M-step at iteration $k\ge 0$ are obtained by repeating the same reasoning but when taking the expected log-likelihood.


Following the same reasoning leading in \ref{app:consN}, it is possible to show that at time $\bar t$ the one-step-ahead MSE of the linear system \eqref{eq:ssidiocorr} reaches a steady-state given by
\[
\bm A = \l(\ba{cc}
\bm\Gamma_1^e&\mbf 0_{m\times 1}\\
\mbf 0_{1\times m}& \sigma_u^2 
\ea
\r).
\]
Now, define the matrices
\begin{align}
\bm B=
\l(
\ba{cc}
\mbf I_m  &\mbf 0_{m\times (n-m)}\\
\mbf 0_{(n-m)\times m}&\bm\Gamma_0^e
\ea
\r),\qquad 
\bm C&=\l(
\ba{cc}
\mbf I_m \phi^{-1/2} &\bm{\mathcal B}_1\phi^{-1/2}\\
\mbf 0_{(n-m)\times m}&\bm{\mathcal B}_0
\ea
\r),\nn
\end{align}
where $\bm C$ is $n\times (m+1)$ and $\bm B$ is $n\times n$, and notice also that $\bm B^{-1}$ is well defined because of Assumption \ref{ass:modelNS}(b).  Then, given the true value of the parameters, $\bm\varphi_n$, for any given $t=\bar t,\ldots, T$, the KF estimator of the states is given by:
\begin{align}
\bm s_{t|t}=\bm s_{t|t-1}+\l(\bm C^\prime \bm B^{-1}\bm C+\bm A^{-1}\r)^{-1}\bm C^\prime \bm B^{-1}(\mbf x_t-\bm C\bm s_{t|t-1}).\label{eq:KFidiocorr}
\end{align}
To prove consistency we need to apply Woodbury formula. However, this is not possible, indeed
\begin{align}
\bm C^\prime\bm C=\phi^{-1}
\l(
\ba{cc}
\mbf I_m  &\bm{\mathcal B}_{1}\\
\bm{\mathcal B}_{1}^\prime&\bm{\mathcal B}_{1}^\prime\bm{\mathcal B}_{1}
\ea
\r)+\l(
\ba{cc}
\mbf 0_{m\times m}  &\mbf 0_{m\times 1}\\
\mbf 0_{1\times m}&\bm{\mathcal B}_{0}^\prime\bm{\mathcal B}_{0}
\ea
\r)=\phi^{-1}\bm{\mathcal C_1}+\bm{\mathcal C_2}, \;\mbox{say}.\label{eq:CCC1C2}
\end{align}
and this is a singular matrix, because $\nu^{(m+1)}(\bm{\mathcal C}_1)= 0$ for all $m\in\mathbb N$ and $\nu^{(j)}(\bm{\mathcal C}_2)= 0$, for $j=2,\ldots, (m+1)$ and all $m\in\mathbb N$. Moreover, notice also that if $\phi=0$ then $\bm C^\prime\bm C$ will not be defined and the KF would have no sense.


However, since the idiosyncratic components are weakly cross-correlated by Assumption \ref{ass:modelNS}(c), it is reasonable to assume that they have a common factor. For simplicity, let us assume that
\beq
\bm\xi_t = \bm\beta w_t, \label{eq:weakfac}
\eeq
with $\E_{\varphi_n}[w_t^2]=1$ and $ \bm\beta=(\beta_1\cdots\beta_m)^\prime$ is such that $m^{-\alpha}\bm\beta^\prime\bm\beta=1$ for some real $\alpha$ and $\vert \beta_i\vert\le M_\beta$ for some positive real $M_\beta$ independent of $i$. In particular,  notice that for $\bm\xi_t$ to be idiosyncratic, thus with $\Delta\bm\xi_t$ satisfying \eqref{eq:divevaldiff2}, we must have $\alpha\in[0,1)$. This is equivalent to saying that the system is driven by a pervasive factors $f_t$ and a local factor $w_t$, which affects weakly only for the first $m$ units. Moreover, since $\bm\beta$ has full column rank, we can write $w_t=m^{-\alpha}\bm\beta^\prime \bm\xi_t$, and therefore
\beq
w_t= m^{-\alpha}\bm\beta^\prime \bm P \bm\beta w_{t-1}+m^{-\alpha}\bm\beta^\prime\mbf e_{1t}.\label{eq:wcorr}
\eeq
Using \eqref{eq:weakfac} and \eqref{eq:wcorr}, the state space formulation in \eqref{eq:ssidiocorr} becomes:
\begin{align}
\l(\ba{c}
\mbf x_{1t}\\
\mbf x_{0t}
\ea
\r)&=
\l(
\ba{cc}
\bm\beta &\bm{\mathcal B}_1\\
\mbf 0_{(n-m)\times m}&\bm{\mathcal B}_0
\ea
\r)\l(\ba{c}
w_{t}\\
f_t
\ea
\r)+
\l(\ba{c}
\bm \nu_{t}\\
\mbf e_{0t}
\ea
\r), 
\nn\\
\l(\ba{c}
w_{t}\\
f_t
\ea
\r)&= \l(\ba{cc}
m^{-\alpha}\bm\beta^\prime \bm P \bm\beta&\mbf 0_{m\times 1}\\
\mbf 0_{1\times m}& 1
\ea
\r)\l(\ba{c}
w_{t-1}\\
f_{t-1}
\ea
\r)+\l(\ba{c}
m^{-\alpha}\bm\beta^\prime\mbf e_{1t}\\
u_t
\ea
\r),
\label{eq:ssidiocorr2}
\end{align}
where the errors have the same distribution as in \eqref{eq:ssidiocorr}.

As a consequence,
\[
\bm A=\l(\ba{cc}
m^{-2\alpha}\bm\beta^\prime\bm\Gamma_1^e\bm\beta&0\\
0& \sigma^2_u
\ea
\r), \qquad 
 \bm C=
\l(
\ba{cc}
\bm\beta\phi^{-1/2}  &\bm{\mathcal B}_1\phi^{-1/2} \\
\mbf 0_{(n-m)\times 1}&\bm{\mathcal B}_0
\ea
\r),
\]
while $\bm B$ is unchanged. The state vector is now $\bm s_t=(w_t\ f_t)^\prime$ and with these new definitions \eqref{eq:KFidiocorr} still holds, while \eqref{eq:CCC1C2} becomes
\begin{align}
\bm C^\prime\bm C=\phi^{-1}
\l(
\ba{cc}
\bm\beta^\prime\bm\beta  &\bm\beta^\prime\bm{\mathcal B}_{1}\\
\bm{\mathcal B}_{1}^\prime\bm\beta&\bm{\mathcal B}_{1}^\prime\bm{\mathcal B}_{1}
\ea
\r)+\l(
\ba{cc}
\mbf 0_{m\times m}  &\mbf 0_{m\times 1}\\
\mbf 0_{1\times m}&\bm{\mathcal B}_{0}^\prime\bm{\mathcal B}_{0}
\ea
\r)=\phi^{-1}\bm{\mathcal C_1}+\bm{\mathcal C_2}, \;\mbox{say}.\label{eq:CCC1C2new}
\end{align}
which is not singular since $\nu^{(2)}(\bm C^\prime\bm C)=\phi^{-1}m^{\alpha}$. 
Moreover, from \citet[Theorem 7]{MK04} and Assumptions  \ref{ass:modelNS}(b) and \ref{ass:modelNS}(c) we can show that $\nu^{(2)}(\bm A)= Mm^{-\alpha}$ for some positive real $M$. 
By following the same steps of the proof of Lemma 14
in \citet{BLqml}, we have
\beq
\l(\bm C^\prime \bm B^{-1}\bm C+\bm A^{-1}\r)^{-1}\bm C^\prime \bm B^{-1}\bm C=\mbf I_2+O(\phi),\label{eq:phi}
\eeq
from which we see that a necessary condition for consistency is $\phi\to0$ as $n\to\infty$. By the same arguments we also have
\beq
\Vert\l(\bm C^\prime \bm B^{-1}\bm C\r)^{-1}\Vert= \phi m^{-\alpha}.\label{eq:phi2}
\eeq
Substituting \eqref{eq:phi} and \eqref{eq:phi2} into \eqref{eq:KFidiocorr}, we have
\begin{align}
\Vert\bm s_{t|t}-\bm s_t\Vert&\le \Vert(\bm C^\prime \bm B^{-1}\bm C)^{-1}\Vert \ \l\Vert\bm C^\prime \bm B^{-1}
\l(\ba{c}
\bm\nu_t\\
\mbf e_{0t}
\ea\r)\r\Vert+ O(\phi)\le K m^{-\alpha}\l\Vert
\l(\ba{c}
\phi^{1/2}\bm\beta^\prime\bm\nu_t\\
\phi^{1/2}\bm{\mathcal B}_1^\prime\bm\nu_t+\phi\bm{\mathcal B}_0^\prime(\bm\Gamma_0^e)^{-1} \mbf e_{0t}
\ea\r)
\r\Vert+ O(\phi)\nn,
\end{align}
for some positive real $K$. Then, 
\begin{align}
&\E_{\varphi_n}\l[\l(m^{-\alpha} \phi^{1/2}\bm\beta^\prime\bm\nu_t\r)^2\r]\le M_\beta^2\phi m^{-2\alpha}\sum_{i=1}^m \E_{\varphi_n}[\nu_{it}^2]= O(\phi^2 m^{1-2\alpha}),\nn\\
&\E_{\varphi_n}\l[\l(m^{-\alpha} \phi^{1/2}\bm{\mathcal B}_1^\prime\bm\nu_t\r)^2\r]\le M_\lambda^2\phi m^{-2\alpha}\sum_{i=1}^m \E_{\varphi_n}[\nu_{it}^2]= O(\phi^2 m^{1-2\alpha}),\nn
\end{align}
since $\E_{\varphi_n}[\nu_{it}\nu_{jt}]=0$ for all $i\ne j$, and where in the second relation we  also used Assumption \ref{ass:modelNS}(a). Moreover,\begin{align}
&\E_{\varphi_n}\l[\l(m^{-\alpha} \phi\bm{\mathcal B}_0^\prime(\bm\Gamma_0^e)^{-1}{\mbf e}_{0t}\r)^2\r]
=m^{-2\alpha} \phi^2\bm{\mathcal B}_0^\prime(\bm\Gamma_0^e)^{-1}\bm{\mathcal B}_0=O(m^{-2\alpha} \phi^2 n),\nn
\end{align}
by Assumptions \ref{ass:modelNS}(b) and \ref{ass:modelNS}(c). Therefore,
\begin{align}
&\vert w_{t|t}- w_t\vert = O_p(\phi m^{-\alpha}\sqrt m)+O(\phi),\qquad\vert f_{t|t}- f_t\vert = O_p(\phi m^{-\alpha} \sqrt n)+O(\phi).\label{eq:wttw}
\end{align}
Consider the simplest case $\alpha=0$, then we need at least $\phi=o(n^{-1/2})$ to achieve convergence. In particular, if we set $\phi=n^{-1}$ we have that $f_{t|t}$ is $\sqrt n$-consistent, whereas $\vert w_{t|t}- w_t\vert=O_p(n^{-1}\sqrt m)$. 

The previous result holds for any $m$ and $n$. However, what we are really interested in is the estimation of the vector $\bm\xi_t=\bm\beta w_t$. 
For given $\bm\beta$, letting $\bm\xi_{t|t}=\bm\beta w_{t|t}$, from \eqref{eq:wttw}, we have
\[
\Vert \bm\xi_{t|t}-\bm \xi_t\Vert = \sqrt{\sum_{i=1}^m \beta_i^2(w_{t|t}- w_t)^2} \le M_\beta \sqrt m \vert w_{t|t}- w_t\vert = O_p(\phi m^{-\alpha} m)+O(\phi\sqrt m).
\]
Hence, when $\alpha=0$, if we still set $\phi=n^{-1}$, we must have $mn^{-1}\to 0$ in order to have consistency. Furthermore, to achieve $\sqrt n$-consistency we would need either $mn^{-1/2}\to 0$ or an even smaller value of $\phi$. 

To conclude, notice that in practice, although the model in \eqref{eq:ssidiocorr} is equivalent to the model in \eqref{eq:ssidiocorr2}, the latter has fewer states but more parameters to estimate and moreover estimation of $\bm\beta$ is not straightforward. In view of this comment the above derivations can just be seen as providing an intuition of the complexity involved by adding $m$ idiosyncratic latent states.

\end{document}